\documentclass[twocolumn]{aastex631}

\usepackage{xcolor}

\newcommand{\ud}{\mathrm{d}}
\newcommand{\msunh}{h^{-1}\mathrm{M}_\odot }
\newcommand{\mpch}{h^{-1}\mathrm{Mpc} }
\newcommand{\kmpch}{h\mathrm{Mpc}^{-1} }
\newcommand{\Rid}{R_{\rm id}}
\newcommand{\Rvir}{R_{\rm vir}}

\usepackage{newtxtext,newtxmath}
\usepackage{graphicx}	
\usepackage{amsmath}
\graphicspath{{./}{figures/}}

\begin{document}
\title{Einasto profile as the halo model solution coupled to the depletion radius}

\author[0000-0002-4260-051X]{Yifeng Zhou}
\affiliation{Department of Astronomy, School of Physics and Astronomy, Shanghai Jiao Tong University, Shanghai 200240, China}
\affiliation{Key Laboratory for Particle Astrophysics and Cosmology (MOE), Shanghai 200240, China}
\affiliation{Shanghai Key Laboratory for Particle Physics and Cosmology, Shanghai 200240, China}

\author[0000-0002-8010-6715]{Jiaxin Han}
\affiliation{Department of Astronomy, School of Physics and Astronomy, Shanghai Jiao Tong University, Shanghai 200240, China}
\affiliation{Key Laboratory for Particle Astrophysics and Cosmology (MOE), Shanghai 200240, China}
\affiliation{Shanghai Key Laboratory for Particle Physics and Cosmology, Shanghai 200240, China}

\correspondingauthor{Jiaxin Han}
\email{yifengzhou@sjtu.edu.cn (YZ), jiaxin.han@sjtu.edu.cn (JH)}



\begin{abstract}

We constrain the halo profiles outside the halo boundaries by solving for the matching profiles required by the halo model. In the halo model framework, the matter distribution in the universe can be decomposed into the spatial distribution of halos convolved with their internal structures. This leads to a set of linear equations in Fourier space which uniquely determines the matching halo profiles for any given halo catalog. 
In this work, we construct three halo catalogs with different boundary definitions, and solve for the matching profiles in each case using measurements of halo-matter and halo-halo power spectra. Our results show that for a given halo field, there is always a set of matching profiles to accurately reconstruct the input statistics of the matter field, even though it might be complex to model the profiles analytically. Comparing the solutions from different halo catalogs, we find their mass distributions inside the inner depletion radii are nearly identical, while they deviate from each other on larger scales, with a larger boundary resulting in a more extended profile. 
For the depletion radius based catalog, the numerical solution agrees well with the Einasto profile. Coupling the Einasto profile with the depletion catalog, the resulting halo model can simultaneously predict the halo-matter power spectra to $10\%$ and matter-matter power spectrum to $5\%$, improving over conventional models in both the interpretability and versatility. {The conditions and limitation of using the Navarro-Frenk-White profile in the halo model are also discussed.} 

\end{abstract}

\section{Introduction} \label{sec:1}

By assuming that all mass is contained within virialized objects called dark matter halos, the mass distribution of the Universe can be decomposed into the spatial distribution of halos, convolved by their internal structures. On large scales, the internal structures of halos become unimportant, and the halo model has been quite successful in tracing the large-scale structure~\citep[see e.g.,][for reviews]{cooray2002halo,Asgari23} of the Universe. On intermediate and small scales, however, an accurate halo model requires detailed modelling of the structures and boundaries of halos.

Conventionally, a halo is defined as a virialized object according to the virial radius, $\Rvir$, as motivated by the spherical collapse model~\citep{gunn1972infall}. Using cosmological simulations, the spherically averaged halo profile out to the virial scale has been found to be well described by some empirical fitting functions such as the Navarro-Frenk-White~\citep[NFW,][]{navarro1995simulations,navarro1996structure,navarro1997universal} or Einasto~\citep{einasto1965construction, Merritt06, Navarro04, Navarro10} profiles. However, the halo model constructed using such profiles truncated at $\Rvir$ does not fully match simulations, especially on the transition scale where halos start to intersect~\citep[e.g.,][]{Garcia19exclusion,HMCODE21,DHM}. 
Some studies have made efforts to address this issue by introducing some global parameters or additional corrections on transition scales~\citep{tinker2005mass,van2013cosmological, HMCODE15, Philcox20EHM}, which further complicates the halo model. Therefore, to build a more concise and accurate halo model, more efforts are needed to quantify the matter distribution around the boundary of a halo as well as the definition of the halo boundary itself. 

Physically, the virial radius is expected to only describe an equilibrium structure, while a growing halo is inevitably surrounded by a non-equilibrium region which extends beyond the virial radius. Significant gravitational influences of a halo on its satellites also start from a radius much larger than the virial radius~\citep{Ludlow09subhalos,strip13Bahe,strip14Behroozi}. Thus, the definition of the halo boundary should be revised to allow for a more extended profile beyond the virial radius. Many recent works have attempted to introduce new boundaries to more physically define a halo. A widely studied candidate is the splashback radius, which is defined at the first apocenter of the orbit of the an infalling particle in a growing halo~\citep{Fillmore84splash,Bertschinger85splash,splashback14Adhikari,DK14,Shi16Splash,Mansfield17splash}. In practice, it is often estimated from the steepest location in the halo density profile~\citep{more2015splashback}. The macroscopic effect of halo growth is to cause a drop in its environmental density. Accordingly, \citet{depletion1} proposed a new halo boundary called the depletion radius which separates the growing part of a halo from the fading environment. According to continuity, the depletion radius can be found at the radius of the maximum mass infall rate.\footnote{\citet{depletion1} introduced two radii called the inner and characteristic depletion radius respectively. Unless explicitly specified, in this work will use the term depletion radius to specifically refer to the inner depletion radius.}
The depletion of the environment due to halo growth is shown to be responsible for the creation of a minimum in the bias profile around a halo, leading to an alternative representation of the depletion radius in the bias domain~\citep{depletion1,depletion2}. These depletion features have also been successfully measured in observations~\citep{li2021outermost,fong2022first}. Some other works have attempted to define the radius of a halo using characteristics in the velocity profile~\citep[e.g.,][]{Cuesta08,Bose21,Pizzardo23infalling1}, phase space structure~\citep{Tomooka20,Aung21} or the mass profile~\citep{Pizzardo23infalling2}. These boundary definitions characterize the halo with different physical motivations, providing us complementary insights to the structure of a dark matter halo. 

Ideally, the halo profile should be self-consistent with its boundary definition so that the profile terminates at the boundary. However, because halos are intrinsically aspherical~\citep{triaxial2002jing,triaxial2006allgood,Mansfield17splash,anisotropy2022wang}, and more importantly, because of ongoing mergers which temporarily extend and distort the domain of a halo, the spherically average density profile can not be crudely truncated at the halo boundary. Instead, a smooth truncation is desired to better describe the halo structure outside a spherical boundary in practice.

A smoothly truncated profile leads to a question: what is the profile outside a given halo boundary? Directly measuring the halo outer profile is difficult in simulations because it is unclear how mass should be partitioned among neighboring halos. To mitigate this problem, some recent works have considered halos and a background density field separately. For example, by partitioning particles around a halo in phase space, the halo profile can be split into orbiting and infalling parts~\citep{DiemerDynamical1, diemer2022dynamics,Garcia3,DynamicalHM24Salazar}
. \citet{ADM2} considered the linear density field as a background and derived the profiles of halos in excess of this background field in Fourier space. Because the density field is a convolution of the halo field with the density profiles of halos, in Fourier space the halo profiles can be solved explicitly and self-consistently from the matter and halo fields. However, because they considered the halo profile in excess of the linear density field, the resulting profiles are not directly applicable to the classical halo model and can have negative values. 

In this work, we will work in the classical halo model framework and study the outer profiles of halos in light of recent developments in the halo boundary definition. Using a similar method to \citet{ADM2}, we will solve for the halo profiles in Fourier space and investigate how the recovered profiles depend on the adopted halo boundary of the input catalog. Transforming the solutions back to real space, we will investigate the analytical properties of the reconstructed profiles, as well as their performances in reproducing additional large scale structure statistics when inserted back to a halo model. Through these analysis, we aim to answer a series of fundamental questions regarding halo profile and halo boundary, including i) For each given halo catalog corresponding to a given halo boundary, is there always a matching set of halo profiles to be used in the halo model? ii) With the matching profile to the halo boundary, how well does the resulting halo model work in predicting additional large scale structure statistics not used to constrain the model? iii) For different boundary choices, is there an optimal boundary to produce the best halo model, and what is the physical implication of such a choice if it exists?

This paper is organized as follows. In Section~\ref{sec:3} we describe the method used for constraining the halo profile. In Section~\ref{sec:2}, we describe the simulation data and introduce three halo catalogs corresponding to different boundary definitions. Section~\ref{sec:4} shows the reconstructed profiles for the three catalogs, and evaluates their performances in predicting the matter-matter power spectrum. We discuss the stability of our method on large scales and the impacts of unresolved mass in Section~\ref{sec:discussion}. Finally, in Section~\ref{sec:6}, we summarize the results of this paper.

\section{Method} \label{sec:3}
In this section we derive the set of linear equations for constraining the halo profiles in Fourier space, and explain how we account for unresolved halos and diffuse matter which distinguishes our model from some other similar methods. 

\subsection{Constraining halo profiles in the halo model framework} \label{sec:3.1}
The cross power spectrum of the matter field and a halo population with mass $M$ and mean number density $n(M)$ can be expressed as
\begin{align}
    P_{\rm hm}(k|M) &= P^{\rm 1h}_{\rm hm}(k|M) + P^{\rm 2h}_{\rm hm}(k|M), \label{eq:Phm-decomp}\\
    P^{\rm 1h}_{\rm hm}(k|M) &= \frac{W(k|M)}{\Bar{\rho}_{\rm m}}, \label{eq:Phm-1halo} \\
    P^{\rm 2h}_{\rm hm}(k|M) &= \frac{1}{\Bar{\rho}_{\rm m}}\int n(m)P^{\rm 2h}_{\rm hh}(k|m,M)W(k|m)\ud m, \label{eq:Phm-2halo} 
\end{align}
where $\bar{\rho}_{\rm m}$ is the mean density of the Universe, $P_{\rm hh}^{\rm 2h}$ is the two halo term of the halo-halo power spectrum between halos of mass $m$ and $M$, adn $W(k|m)$ is the halo density profile in Fourier space. The relations of $k$-space profile $W(k|M)$ and real-space profile $\rho(r|M)$ are given by
\begin{align}
    W(k) &= \int^{\infty}_{0}\rho(r|M)\frac{\sin(kr)}{kr}4\pi r^2 \ud r, \\
    \rho(k) &= \frac{1}{(2\pi)^3 }\int^{\infty}_{0}\rho(r|M)\frac{\sin(kr)}{kr}4\pi k^2 \ud k.
\end{align}
In this case, $W(k)$ converges to the integrated mass $M_{\rm int}$ of the density profile when $k$ goes to 0. Equations~\eqref{eq:Phm-decomp} to \eqref{eq:Phm-2halo} show clearly that the halo-matter power spectrum is the linear combination of the halo profiles, which implies that the halo profiles can be completely solved with enough independent constraints.

We now proceed to divide halos into $l$ mass bins to establish such constraints. Considering the halo-matter power spectrum of the $i$th mass bin at $k$, Equation~\eqref{eq:Phm-decomp} can be rewritten as,
\begin{align}
    P_{\rm hm}(k|m_i) &= \frac{W(k|m_i)}{\Bar{\rho}_{\rm m}} \nonumber \\
    &+\frac{1}{\Bar{\rho}_{\rm m}}\sum^l_{j=1}n(m_j)P^{\rm 2h}_{\rm hh}(k|m_j,m_i)W(k|m_j).
    \label{eq:constraint}
\end{align}
Here the integral over mass has been replaced with the summation. Equation~\eqref{eq:constraint} contains $l$ unknown variables $W(k|m_j)$ at each $k$. By combining the halo-matter power spectra from $l$ mass bins, a set of solution for $W(k|m_j) (j=1..l)$ can be completely determined.

To organize the equations into a compact form, we define the following vectors 
\begin{align}
     \textbf{\textit{h}} &= \left(
\begin{array}{cccc}
P_{\rm hm}(k|m_1) \\
P_{\rm hm}(k|m_2) \\
\vdots \\
P_{\rm hm}(k|m_l) \\
\end{array}
\right), \nonumber \\
    \textbf{\textit{w}} &= \left(
\begin{array}{cccc}
W(k|m_1) \\
W(k|m_2) \\
\vdots \\
W(k|m_l) \\
\end{array}
\right).
\end{align}
and tensors
\begin{align}
     \textbf{H} &= \left(
\begin{array}{cccc}
P_{\rm hh}(k|m_1, m_1) & P_{\rm hh}(k|m_1, m_2) & \cdots & P_{\rm hh}(k|m_1, m_l) \\
P_{\rm hh}(k|m_2, m_1) & P_{\rm hh}(k|m_2, m_2) & \cdots & P_{\rm hh}(k|m_2, m_l) \\
\vdots \\
P_{\rm hh}(k|m_l, m_1) & P_{\rm hh}(k|m_l, m_2) & \cdots & P_{\rm hh}(k|m_l, m_l) \\
\end{array}
\right), \nonumber \\
\textbf{N} &= \left(
\begin{array}{cccc}
n(m_1)& \cdots & \cdots & 0 \\
0 & n(m_2) & \cdots & 0 \\
\vdots&\vdots&\vdots&\vdots \\
0 & \cdots & \cdots & n(m_l) \\
\end{array}
\right)
\end{align}
So that, the constraints on the halo profiles can be rewritten as
\begin{equation}
    \textbf{\textit{h}} = \frac{1}{\bar{\rho}_{\rm m}}(\textbf{I}+\textbf{HN})\textbf{\textit{w}}.
    \label{eq:equations_SHM}
\end{equation}
{where $\textbf{I}$ is the identity matrix.} 


\subsection{Accounting for unresolved halos and diffuse matter} \label{sec:3.3}
Theoretically, the halo mass bins in Equation~\eqref{eq:equations_SHM} should cover the complete mass spectrum of halos. However, due to free streaming of cold dark matter particles\citep[e.g.,][]{Green05,Profumo06,Schneider13}, and due to the finite resolution of numerical simulations, halos can only be resolved and modeled down to a certain mass limit. It is thus necessary to introduce an unresolved mass component to account for the contribution from unresolved halos as well as from some potential diffuse matter, so that Equation~\eqref{eq:equations_SHM} becomes
\begin{equation}
    \textbf{\textit{h}} = \frac{1}{\bar{\rho}_{\rm m}}(\textbf{I}+\textbf{HN})\textbf{\textit{w}}+\textbf{\textit{c}}.
    \label{eq:equations_DHM}
\end{equation} Here we have defined a vector 
\begin{equation}
    \textbf{\textit{c}} = \left(
\begin{array}{cccc}
P_{\rm hm}^{\rm u}(k|m_1) \\
P_{\rm hm}^{\rm u}(k|m_2) \\
\vdots \\
P_{\rm hm}^{\rm u}(k|m_n) \\
\end{array}
\right),
\end{equation}
where $P_{\rm hm}^{\rm u}$ is the cross power spectrum between halo and unresolved mass.

\citet{DHM} modeled the unresolved component $\xi^{\rm u}_{\rm hm}(r|m)$ in real space by approximating the unresolved halos as mass points distributed outside the exclusion radius, so that $\xi^{\rm u}_{\rm hm}(r|m)$ can be expressed as a universal halo-halo correlation $\hat{\xi}_{\rm hh}(r)$ multiplied by the effective bias of the unresolved component, $b_{\rm unr}$, with a truncation due to halo exclusion. 
{We will model the unresolved halo-matter correlation function following \citet{DHM}'s form and convert it to $P_{\rm hm}^{\rm u}$ in Fourier space.}
Equation~\eqref{eq:equations_DHM} forms the basis for our solution to the halo density profile. {With $\textbf{\textit{h}}$, $\textbf{H}$, $\textbf{N}$ measured from the simulation data and a theoretical model for $\textbf{\textit{c}}$, $\textbf{\textit{w}}$ can be solved directly from the above equation, independently for each $k$ mode. The measurements of $\textbf{\textit{h}}$, $\textbf{H}$ are detailed in Section~\ref{sec:measuring_P}, and the modeling of  $\textbf{\textit{c}}$ is detailed in Appendix~\ref{app:unresolved_term}.} We also compare the solutions with and without the unresolved term in Section~\ref{sec:discussion}, and find that the unresolved term impact the mass distribution in the halo outskirts, especially for low mass halos.

\section{Simulation and halo samples} \label{sec:2}
\subsection{Simulation data} \label{sec:2.1}
We use a Lambda Cold Dark Matter simulation, which is one of the CosmicGrowth simulations \citep{jing2019cosmicgrowth} run with a P$^3$M code, to extract the data of halo and matter fields. The simulation was run in a box of side length 600$\mpch$ containing $3072^3$ particles with cosmological parameters $\Omega_{\rm m}=0.268$, $\Omega_{\Lambda}=0.732$, $h=0.704$, $n_{\rm s}=0.968$, and $\sigma_8 = 0.830$.

The candidate halos are identified by the the Friends-of-Friends (FoF) algorithm with a standard linking parameter of 0.2, and then processed with HBT+ \citep{han2012resolving,HBT} to identify subhalos. We define the virial mass of a halo as the mass enclosed in a sphere with a virial density according to the spherical collapse model \citep{sphericalcollapse2}, and the corresponding radius is the virial radius, $\Rvir$. The depletion radius $R_{\rm id}$, is defined as the location of the maximum mass inflow rate. In this work, we estimate the depletion radius using the scaling relation $\Rid$=2.1$\Rvir$ for individual halos~\citep{depletion1,depletion2}.

We construct an original sample with about $2\times10^6$ distinct halos by selecting candidate objects within a mass range of $11.50<{\rm log}_{10}[m/(h^{-1}{\rm M}_{\odot})]<15.35$ at z=0. These halos are divided into seven logarithmic mass bins with equal spacing. To investigate how the definitions of the exclusion radius affect the halo outer profiles, halos are further selected according to the exclusion criteria, resulting in three halo catalogs. We put more details about the cleaning in Section~\ref{sec:2.2}. 

\subsection{Halo catalogs with different exclusion criteria} \label{sec:2.2}
\citet{DHM} has demonstrated that the matter field can be decomposed into some self-similar halo distributions convolved by the Einasto profiles when halos are selected according to the $\Rid$. Considering the physical picture that the $\Rid$ separate the growing halo and the depleting environment~\citep{depletion1,depletion2}, it is natural to choose the $\Rid$ as the exclusion radius of halos. This choice is also supported by the fact that $\Rid$ was found to coincide with the optimal exclusion radius in a flexible halo model \citep{Garcia2}. 

Nevertheless, as we will show in Section~\ref{sec:4} below, it is still possible to build accurate halo models using other halo catalogs than the depletion catalog, at least for reproducing the halo-matter correlation as long as appropriate halo profiles are chosen.

In this work, we mainly focus on the halo boundary defined as the depletion radius. Meanwhile, we also want to explore how the corresponding solutions vary when some other boundaries are adopted which can be either smaller or larger than the depletion radius. To investigate these different solutions, we construct three halo catalogs by selecting halos according to different exclusion radii from the original catalog. These catalogs are named as
\begin{itemize}
    \item $\Rvir$ catalog: The exclusion radius $R_{\rm ex}$ is defined as the virial radius; 
    \item $\Rid$ catalog: The exclusion radius $R_{\rm ex}$ is defined as the inner depletion radius; 
    \item 3$\Rvir$ catalog: The exclusion radius $R_{\rm ex}$ is defined as three times the virial radius. 
\end{itemize}
The $\Rvir$ boundary is adopted as it is the classical boundary of a halo, while also representing the case for a smaller boundary than the $\Rid$. On the other hand, the $3\Rvir$ is used simply to represent a boundary choice larger than the depletion radius, with no particular physical motivations.
For each catalog, we remove halos which intersect with a more massive neighbor on their exclusion radii. Specifically, {if the distance between a halo pair is smaller than the sum of their exclusion radii, $r_{\rm ex} \equiv R_{\rm ex}(m_1)+R_{\rm ex}(m_2)$, we remove the smaller one from the catalog.} 
The remaining halos form a cleaned catalog with statistics such as the halo mass function and halo-halo correlation different from those in the original catalog. In practice, FoF halos hardly overlap in the virial region because the linking parameter is optimized for disentangling halos according to the virial radius. Thus, the resulting statistics of the original and $\Rvir$ catalogs are nearly identical. 

\begin{figure*}
	\includegraphics[width=\textwidth]{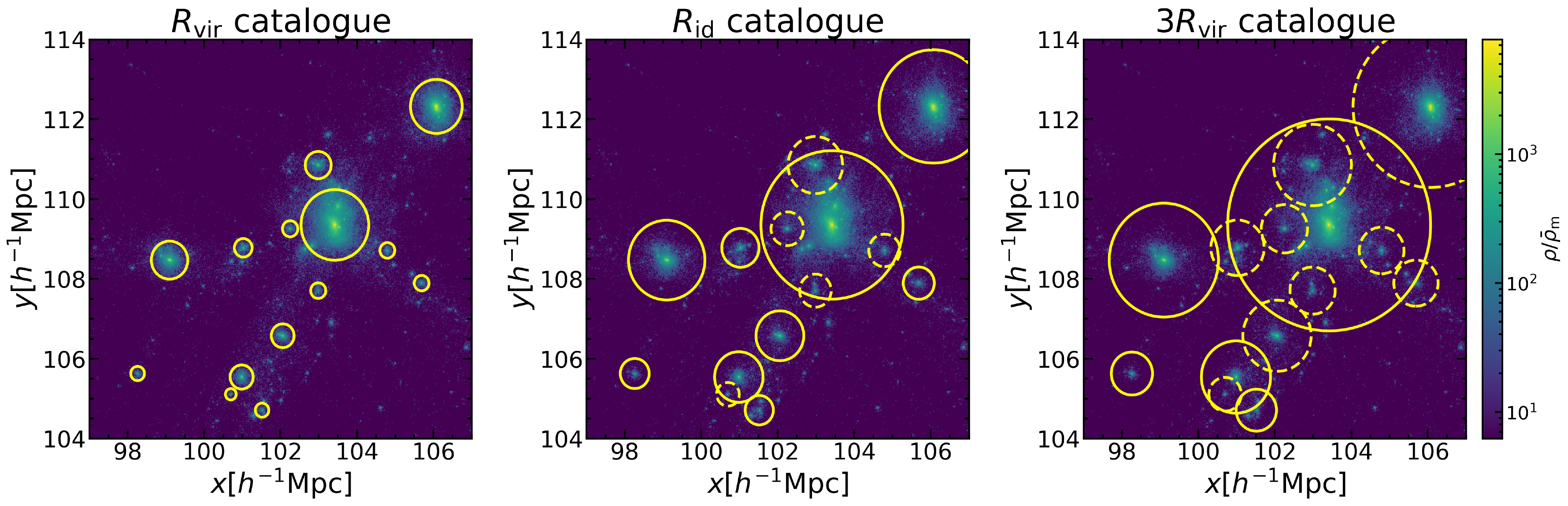}
    \caption{The halo distributions in three catalogs. The projected map of a small region in the simulation box with a thickness of 3$h^{-1}{\rm Mpc}$ is shown, with halos marked by solid circles according to different boundary definitions from left to right. The dashed circles mark removed halos in each catalog according to the exclusion criteria. There are several solid circles overlapping because of the projection effect.}
    \label{fig:halo_distributions}
\end{figure*}

Figure~\ref{fig:halo_distributions} illustrates how halos are distributed in different catalogs. In the left panel, all candidate halos are ``isolated" and contained in the $\Rvir$ catalog since they do not overlap in their virial regions. As the exclusion radius increases in the right panels, more and more halos are removed from the catalog, and halos become more sparse. Correspondingly, it is natural to expect a steeper outer profile when halos are selected with a smaller exclusion criteria, and vice versa, to avoid double counting masses from halos.

\subsection{Measuring the power spectra}
\label{sec:measuring_P}
To solve for the halo profile using Equation~\eqref{eq:equations_DHM}, the power spectrum vectors $\textbf{\textit{h}}$, $\textbf{H}$ should be measured from the simulation. In this work, the auto and cross-power spectra are computed from the three-dimensional matter density field and halo fields in seven mass bins using \textsc{pylians} \citep{Pylians}, a method based on fast Fourier transform (FFT). We use the cloud-in-cell (CIC) scheme to assign particles to the mesh. The simulation box with side length $L=600\mpch$ is divided into $N^3=1024^3$ voxels so that the smallest and largest $k$ modes accessible should be determined as $k_{\rm min}=2\pi/L$ and $k_{\rm max}=\pi N/L$ where $k_{\rm max}$ is estimated as the Nyquist frequency. 

For a Gaussian field, the statistical error on the power spectra can be estimated as \citep{P_error1, P_error2} 
\begin{align}
    \Delta P_{\rm aa} &= \left(\frac{2}{N_{\rm mode}}\right)^{1/2}\left[P_{\rm aa}+P^{\rm shot}_{\rm a}\right], \label{eq:P_errors} \\ 
    \Delta P_{\rm ab} &= \left(\frac{2}{N_{\rm mode}}\right)^{1/2}\sqrt{\frac{P^2_{\rm ab}+(P_{\rm aa}+P^{\rm shot}_{\rm a})(P_{\rm bb}+P^{\rm shot}_{\rm b})}{2}}, \label{eq:error_cross}
\end{align}
where $P_{\rm aa}$ and $P_{\rm ab}$ are the auto and cross power spectra. $P^{\rm shot}$ is the shot noise. $N_{\rm mode}(k)\simeq 2\pi/(k\sqrt{\Delta k V})$ is the number of $k$ modes in a bins centered at $k$, $\Delta k$ is the bin width,  and $V$ is the volume of the box. 

In practice, the raw power spectrum $P_{\rm r}(k)$ obtained from FFT is a smoothed version of the true power spectrum due to the window function effect from mass assignment scheme, and suffers from the aliasing effect due to discrete sampling of the field \citep{aliasing1, aliasing2}. To correct for these effects, we also compute the power spectrum from the Fourier transform of the correlation function,
\begin{equation}
    P'(k) = \int_{r_{\rm min}}^{r_{\rm max}} \xi (r)\frac{\sin(kr)}{kr} 4\pi r^2 \ud r,
    \label{eq:P_correction}
\end{equation}
where $r_{\rm min}=0.02h^{-1}{\rm Mpc}$ and $r_{\rm max}=90h^{-1}{\rm Mpc}$. The correlation function $\xi(r)$ is directly measured from the simulation for $0.02\sim20h^{-1}{\rm Mpc}$ and extrapolated to $90h^{-1}{\rm Mpc}$ using the linear correlation function computed with \textsc{colossus} \citep{diemer2018colossus}. With $r_{\rm min}$ much smaller than the grid size of the FFT, the Fourier transform of the correlation function is closer to the true power spectrum on small scale, but may lose some large-scale information due to our limited $r_{\rm max}$ in the transform. We thus combine the two to give our final measurement of the power spectrum as 
\begin{equation}
    P(k) = \left\{
    \begin{array}{rcl}
         P_{\rm r}(k)      & & k<0.5h{\rm Mpc}^{-1} \\
         P'(k)  & & k\geq 0.5h{\rm Mpc}^{-1}
    \end{array} \right.
    \label{eq:corrected_power_spectrum}
\end{equation}
{where $P_{\rm r}(k)$ is the raw power spectrum measure from the simulation using using \textsc{pylians} \citep{Pylians}, $P'(k)$ is the correction from Equation~\eqref{eq:P_correction}, and $P(k)$ is the corrected power spectrum.}

\begin{figure}
	\includegraphics[width=\columnwidth]{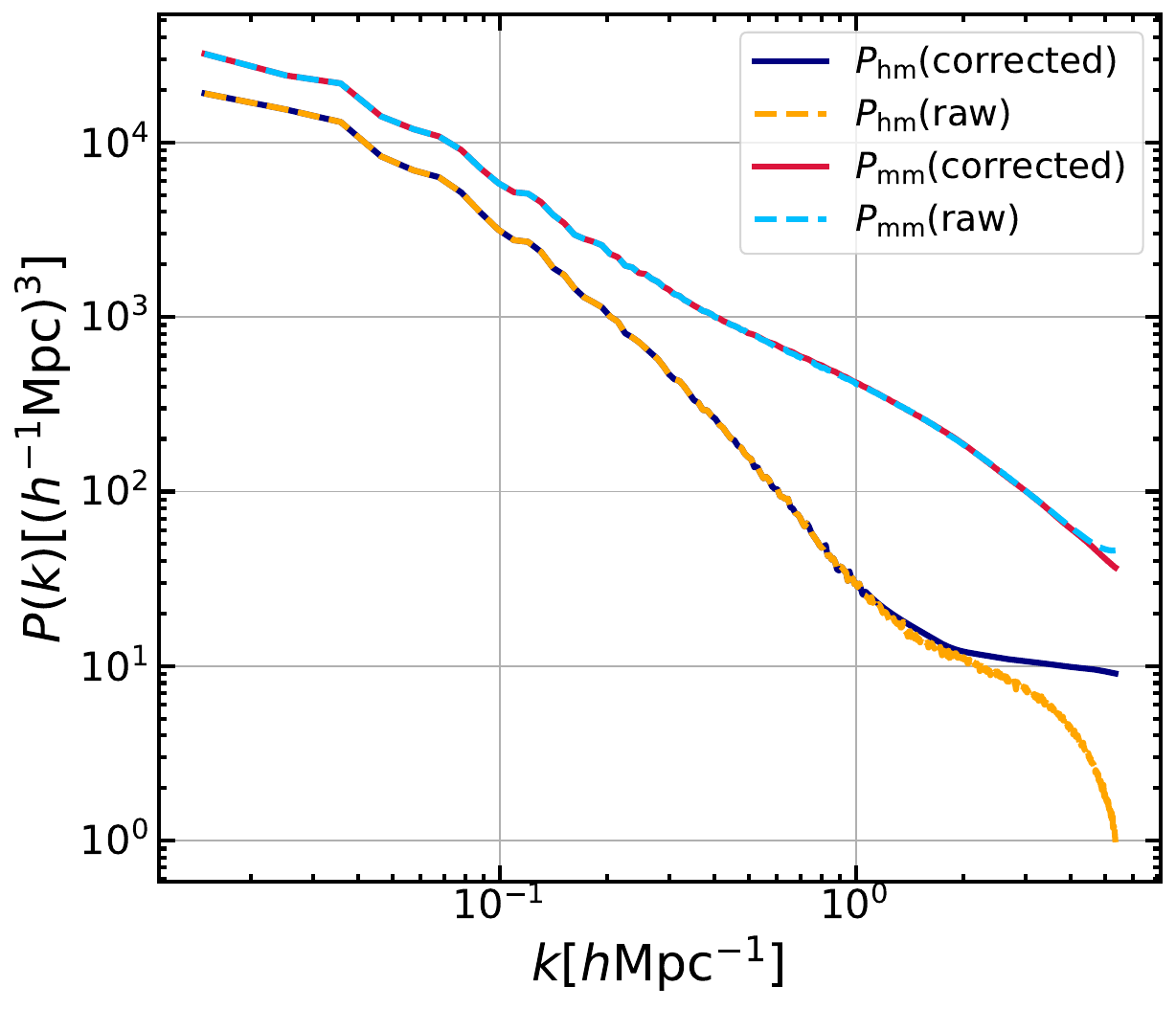}
    \caption{The raw power spectra measured using \textsc{pylians} \citep{Pylians} (dotted curves) and the corrected power spectra from Equation~\eqref{eq:corrected_power_spectrum} (solid curves). 
    $P_{\rm hm}$ is the cross power spectrum of the matter field and the halo population with mass $10^{11.50}<(m/h^{-1}{\rm M}_{\odot})<10^{12.05}$.}
    \label{fig:P_corrections}
\end{figure}

Figure~\ref{fig:P_corrections} shows the raw and corrected power spectra for $k<k_{\rm max}$. We find that the FFT estimation of the halo-matter power spectrum is significantly biased for $k>1\kmpch$. This is mostly due to the window function effect which smoothes out the halo-matter correlation on small scale. After correction, the power spectrum is higher. For the matter-matter power spectrum,  the raw power spectrum is affected by numerical effects near the Nyquist frequency. We do not correct the halo-halo power spectra because the exclusion scale $k_{\rm ex} = 1/r_{\rm ex}$ is smaller than $1\kmpch$ for most halo pairs, beyond which the halo-halo power spectrum is unimportant as we will discuss in Section \ref{sec:discussion}.

To concentrate on the transition region we are interested in, in the following we will focus on the power spectra in the range of $0.1<k<3\kmpch$, which is a narrower range than $k_{\rm min}<k<k_{\rm max}$. This $k$ range corresponds to a radial range from several to about sixty $\mpch$, containing information about halo profiles in the transition region. 

\begin{figure*}
	\includegraphics[width=\textwidth]{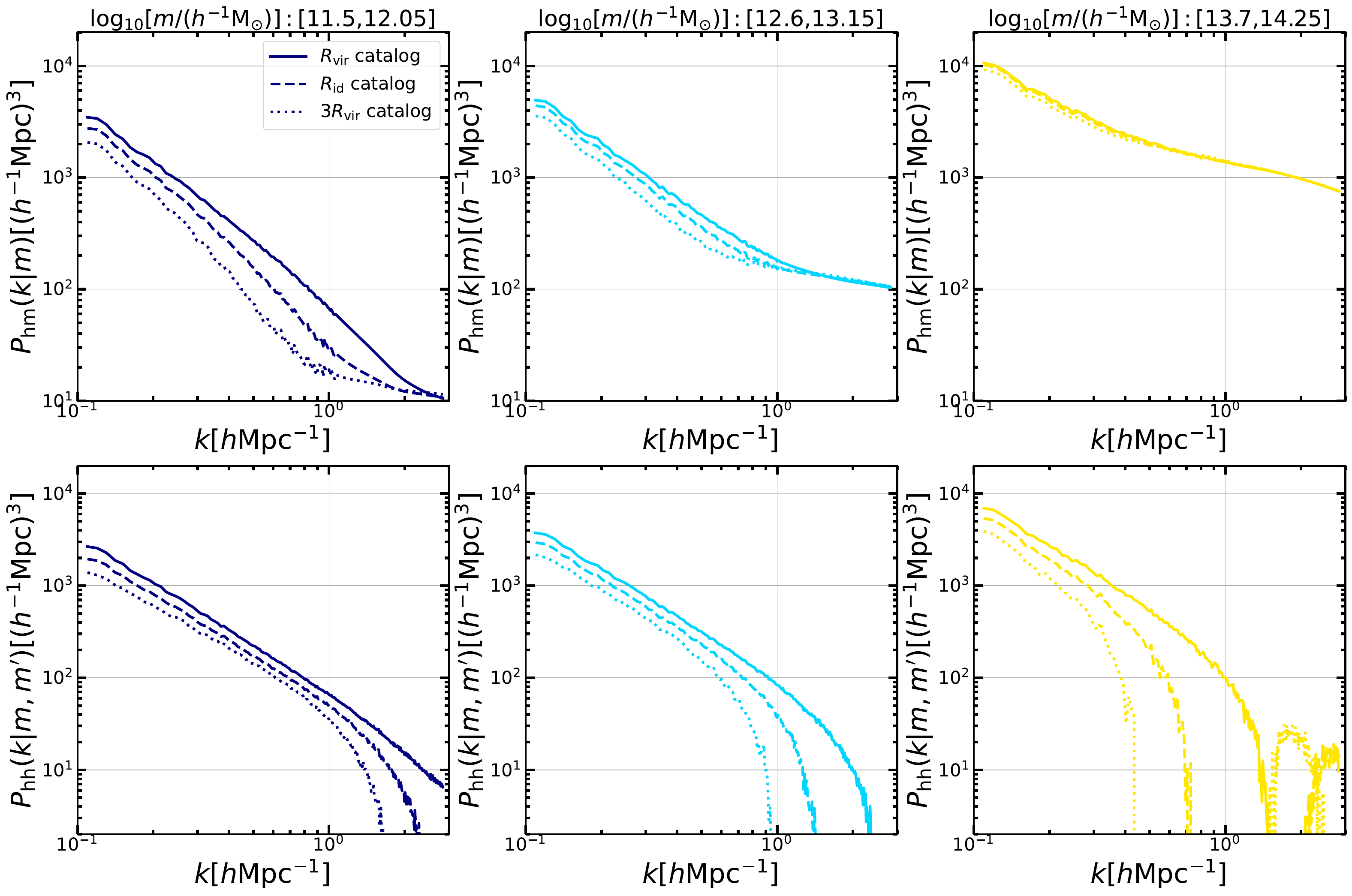}
    \caption{Comparison of population statistics from different 
    halo catalogs. \textit{Top}: the halo-matter power spectra $P_{\rm hm}(k|m)$ for logarithmic mass bins $[11.50, 12.05]$, $[12.60, 13.15]$, and $[13.70,14.25]$. \textit{Bottom}: halo-halo power spectra $P_{\rm hh}(k|m,m^{\prime})$. $m^{\prime}$ is in the logarithmic mass bin $[11.50, 12.05]$ and $m$ are in the same mass bins as top plots.}
    \label{fig:statistics_comps}
\end{figure*}

\citet{Garcia19exclusion} and \citet{DHM} have illustrated how the exclusion scheme impacts the statistics of halo populations in real space. Here, we show some statistics of different halo catalogs in Fourier space. Figure~\ref{fig:statistics_comps} shows the halo-matter power spectra $P_{\rm hm}$s and the halo-halo power spectra $P_{\rm hh}$s of three mass bins. For a given halo catalog, with increasing halo mass $P_{\rm hm}$ and $P_{\rm hh}$ are higher and $P_{\rm hh}$ truncates at a larger scale. This is because the halo bias $b$ and the exclusion scale $r_{\rm ex}$ are increasing functions with respect to halo mass $m$. Comparing the measurements from different catalogs for a given mass bin, $P_{\rm hm}$ and $P_{\rm hh}$ are lower with increasing exclusion radius because halos are less clustered in a catalog with larger exclusion radii. On small scales, $P_{\rm hm}$s converge because the exclusion scheme hardly affects the small-scale matter distribution around halo centers. $P_{\rm hh}$ truncates at larger scales with increasing exclusion radius. Across mass bins, we find the difference between $P_{\rm hm}$s of different catalogs decreases with increasing halo mass, since the exclusion scheme significantly affects the statistics of low-mass halos but the impacts become weak for high-mass halos. 

\section{Numerical solutions for the halo density profiles} \label{sec:4}
\subsection{Matching halo profiles in Fourier space} \label{sec:4.1}
We measure the halo mass function, the halo-halo power spectra, and the halo-matter power spectra from the simulation, and then solve Equation~\eqref{eq:equations_DHM} to obtain the halo profiles. As the large-scale statistics of halos depend on the exclusion criterion, the solution to Equation~\eqref{eq:equations_DHM} varies with halo catalogs. We refer to these solutions as the matching profiles to each halo catalog. 

\begin{figure*}
	\includegraphics[width=\textwidth]{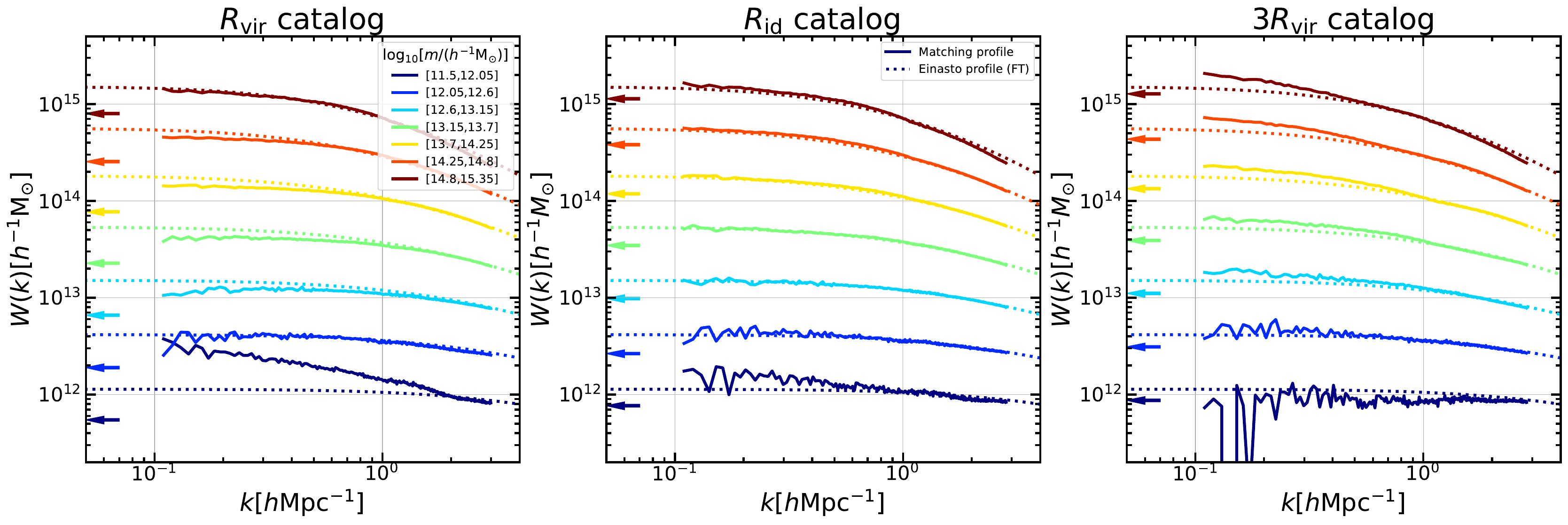}
    \caption{The matching profiles in Fourier space for the three catalogs. Solid curves are the profiles solved from Equation~\ref{eq:equations_DHM}, {while the dotted curves are the Einasto profiles fitted  in the real space and transformed into Fourier space} (identical across three panels). Different colors correspond to different mass bins. {For each profile, the arrow on the left marks the total mass enclosed in its exclusion radius, i.e., $M(<R_{\rm vir})$, $M(<R_{\rm id})$ and $M({\rm <3\Rvir}$).}}
    \label{fig:profiles_kspace}
\end{figure*}

Figure~\ref{fig:profiles_kspace} shows the matching profiles in Fourier space. For the depletion catalog, \citet{DHM} found that the Einasto profile works well for constructing a halo model in real space. We thus fit the halo profiles in the depletion catalog within $\Rid$ using the Einasto formula in real space and invert them into Fourier space as references.  

On small scales of $k>1\kmpch$, the matching profiles follow the corresponding Einasto profiles well in all catalogs for halos with mass $M>10^{12.05}h^{-1}M_{\odot}$. This is consistent with the fact that the inner profiles of halos are barely affected by the exclusion criteria.

On scales $k<1\kmpch$, the profiles are more complicated. As $k\rightarrow0$, $W(k)$ approaches the integrated mass of the density profile. It is thus expected that this asymptotic amplitude of $W(k)$ will increase as a larger exclusion radius is adopted, which is indeed the case in Figure~\ref{fig:profiles_kspace} except for the lowest mass bin. In the $\Rvir$ catalog, the optimal profiles are generally lower than the Einasto fits, while they are higher than the references in the 3$\Rvir$ catalog. By contrast, the profiles in the $\Rid$ catalog still follow the Einasto model well. For each profile, we also show the enclosed mass up to the corresponding halo radius, $M(<r_{\rm ex})$, in each catalog, marked by arrows on the left of panels of Figure~\ref{fig:profiles_kspace}. As expected, $M(<r_{\rm ex})$ is lower than the integrated mass of the same profile as the profiles all extend beyond the exclusion radii.

In the lowest-mass bin, however, the matching profiles no longer follow the above expectations. In the $\Rvir$ catalog, the matching profile of the lowest-mass bin is significantly higher than the Einasto fit and has not converged to a constant towards lower $k$. This can be largely attributed to improper modeling of the unresolved term (see Appendix~\ref{app:unresolved_term}) in this catalog when solving for the matching profile. The lowest mass bin has masses closest to the unresolved halos and are thus the most degenerate with the unresolved component. We discuss the influence of the unresolved component further in section~\ref{sec:discussion}.

Another notable feature is that there are some fluctuations on scales $k<1\kmpch$, especially in low-mass bins. According to Equation~\eqref{eq:P_errors}, the measurements of the power spectrum on larger scales have higher uncertainties. These uncertainties can be propagated to the solution via Equation~\eqref{eq:equations_DHM}. Low mass halos have a lower $W(k)$, resulting in a much larger relative fluctuation. 


We verify that the solutions can accurately reproduce the halo-matter power spectra by inserting the matching profiles back into Equation~\eqref{eq:equations_DHM}, and found residuals less than $10^{-12}$. 
Thus we conclude that, for a given halo catalog, there is indeed a set of matching profiles to accurately reconstruct the mass distribution around halos, at least in the halo-matter correlation. 

\subsection{Matching profiles in real space} \label{sec:4.2}
To gain more physical insights into halo profiles, we transform the matching profiles into real space. As our Fourier space solutions only cover the scales $0.1<k<3\kmpch$, extrapolations to both the high-$k$ and low-$k$ ends are needed before we can transform them to real space. According to the behaviors discussed above, we extrapolate the profiles using the Einasto fits on scales $k>3\kmpch$, and using constant values of $W(k=0.1\kmpch)$ for $k<0.1\kmpch$. The inverse Fourier transform (IFT) is evaluated over the $k$ range of $0<k<10^3 \kmpch$. Note that there are some potential numerical effects in the IFT. For instance, the IFT results can be affected by the missing high-$k$ modes due to the finite $k$ range used, resulting in noises at high frequency. In addition, the interpolation 
of k-space profiles can lead to false signals in the IFT results. To reduce noises caused by these effects, the final results are further smoothed using the Savitzky–Golay algorithm. We also invert the $k$-space Einasto profiles to real space by integrating over the same $k$ range to avoid numerical effects in the comparison.  

\begin{figure*}
	\includegraphics[width=\textwidth]{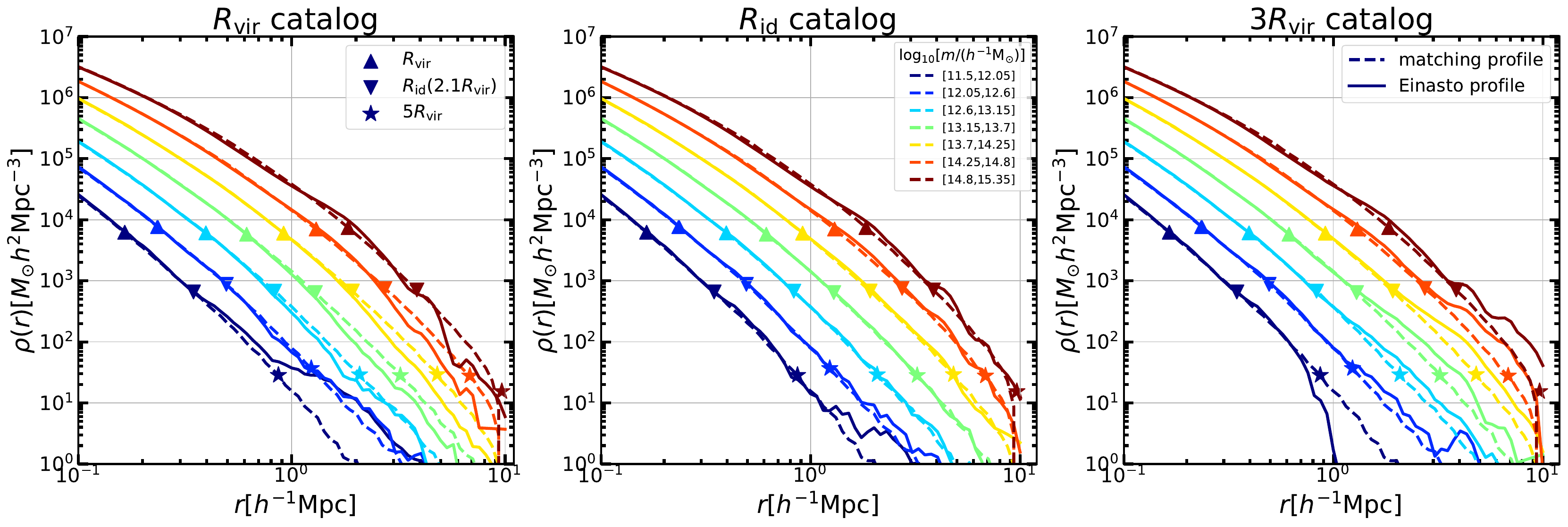}
    \caption{The matching profiles for the three catalogs in real space. In each panel,  the solid and dashed curves are the matching profiles and the reference Einasto profiles in real space, respectively. The upper and lower triangles mark the locations of $\Rvir$ and $\Rid$ respectively. For reference, we also mark the locations of 5$\Rvir$ with stars, within which the $\Rid$ matching profiles agree well with the Einasto profile}.
    \label{fig:profiles_rspace}
\end{figure*}

Figure~\ref{fig:profiles_rspace} shows the matching profiles in real space. For the $\Rid$ catalog, the profiles follow the Einasto profiles within 5$\Rvir$. On scales $r>5\Rvir$, the matching profiles deviate from the Einasto profiles, and are affected by numerical perturbations. 
In the $\Rvir$ catalog, the matching profiles follow the Einasto profiles within $\Rid$ for all mass bins. Outside $\Rid$ they drop faster than the Einasto profiles and are also perturbed due to numerical effects. The exception comes from the lowest-mass bin whose profiles are higher than the Einasto profiles on scales $r>\Rid$. Similarly, the matching profiles to the 3$\Rvir$ catalog follow the Einasto profiles within $\Rid$, but become higher outside $\Rid$ except for the lowest mass bin. These behavior are all consistent with our previous discussions in frequency space. 

Combining the three panels, it is very interesting to notice that the matching profiles inside $\Rid$ are nearly identical across the catalogs, while the outer profiles become more extended with increasing exclusion radii. 


In the $\Rid$ catalog, the profiles follow the Einasto fits well out to a very large scale. This result agrees with the findings of \citet{DHM}, in which the halo-matter correlation functions are accurately modeled when coupling Einasto profiles to the $\Rid$ catalog. In this work, we directly solve the profile matched with the $\Rid$ catalog and find that the optimal profile can be analytically modeled with the Einasto formula, which automatically explains the choice of the halo profile in the Depletion-radius-based Halo Model (here after DHM) of \citet{DHM}. The deviations outside the 5$\Rvir$ shown in Figure~\ref{fig:profiles_rspace} do not challenge this conclusion since the profile on very large scales is unimportant compared with the 2-halo term. 

\section{Towards an explicitly verifiable and versatile halo model for the large scale structure} \label{sec:4.3}




%

The numerical solutions in section~\ref{sec:4} illustrate that a matching halo profile exists for each halo catalog in accurately reproducing the halo-matter power spectrum. The remaining question is then whether and how these solutions may be used to construct accurate analytical halo models. To this end, it is desirable to have analytical prescriptions for each of the model components involved, including the halo profile, halo-halo correlation, halo mass function and the unresolved component which were mostly handled numerically rather than analytically in section~\ref{sec:4}. Establishing analytical prescriptions for these components is not always straightforward due to the complexity and non-universality of the components in some catalogs.


For the $\Rid$ catalog, however, we have showed that the Einasto profile is a good analytical model for the matching profile. In fact, such a profile has already been adopted in \citet{DHM} when constructing the real-space halo model based on the depletion catalog, along with analytical recipes for all other components. It is thus straight-forward to translate the model of \citet{DHM} to Fourier space for the prediction of the power spectrum. Now, we proceed to check how well this model works. 



\begin{figure*}
	\includegraphics[width=\textwidth]{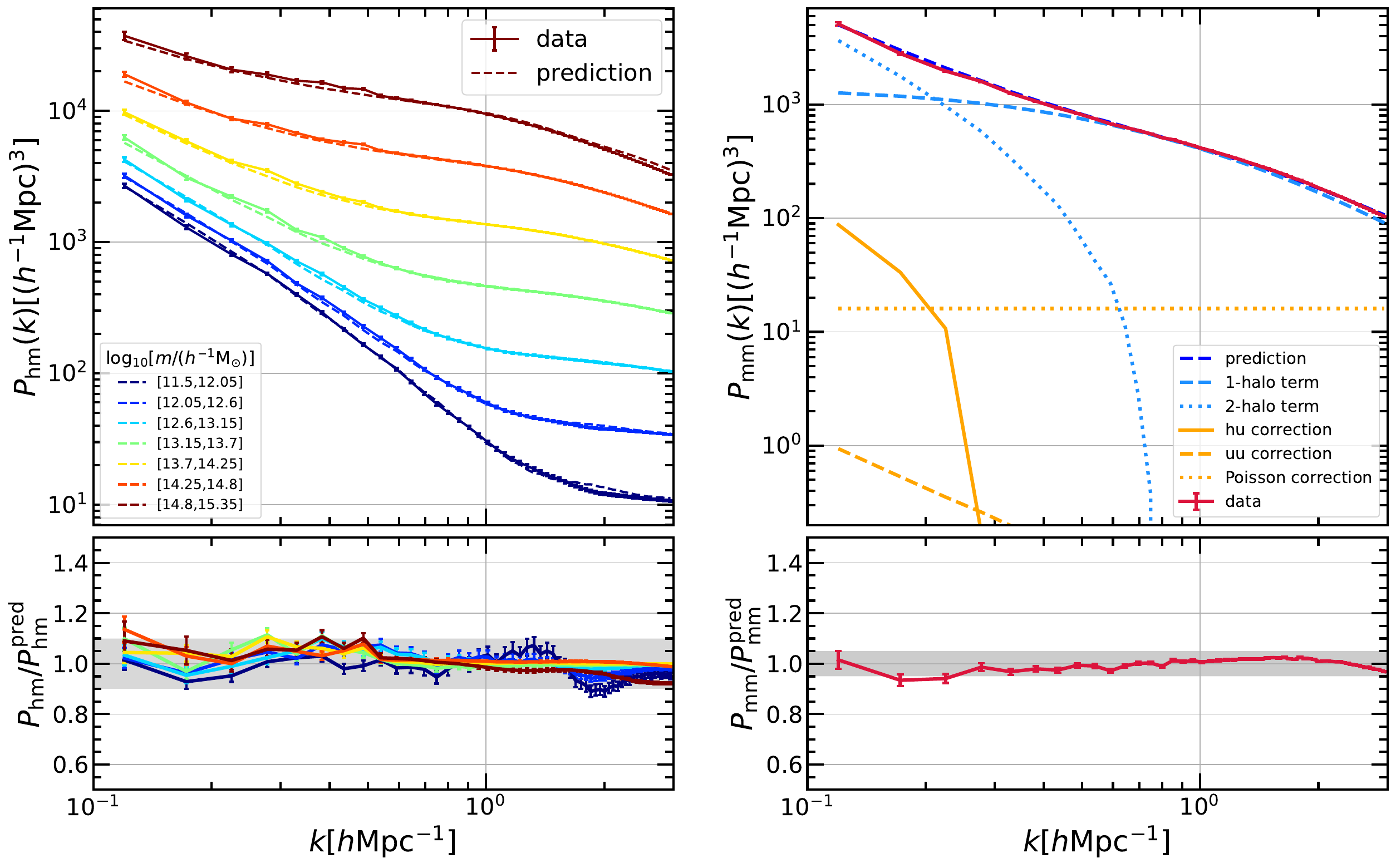}
    \caption{\textit{Left}:The halo-matter power spectra from the simulation and the DHM. The errors are estimated using Equation~\eqref{eq:error_cross}. The bottom panel shows the ratios between the simulated and model power spectra. The shaded areas mark the $10\%$ difference level. \textit{Right}: The matter-matter power spectrum from the simulation and the DHM. The red curve with error bars are simulated power spectrum. The model prediction is decomposed into different components including the resolved 1-halo and 2-halo terms, the halo-unresolved (hu), unresolve-unresolved (uu) and the Poisson terms as detailed in Appendix~\ref{app:1}. The bottom panel shows the ratios between the simulated and model power spectra. The shaded areas mark the $5\%$ difference level.}
    \label{fig:Precons}
\end{figure*}

To construct the model, the profile parameters and the statistics of halos are generated using the fitting formula in \citet{DHM}. The analytical modeling of the ingredients are specified in Appendix~\ref{app:ingredients}. The Einasto profiles, the halo-halo correlations, and the unresolved halo-matter correlations are then converted to Fourier space and inserted into Equation~\eqref{eq:equations_DHM} to predict the halo-matter power spectra. The predictions are shown in the left panel of Figure~\ref{fig:Precons}. The error bars show the standard errors estimated using Equation~\eqref{eq:error_cross}. We find that the DHM performs well in predicting the halo-matter power spectra, achieving about 10$\%$ accuracy (within statistical errors). This performance agrees with that of DHM in real space.  

Ideally, a perfect halo model can fully match the entire density field of the universe, thus capable of predicting multiple statistics of the halo and matter field simultaneously. In Figure~\ref{fig:Precons} we show that the same DHM can also accurately predict the total matter power spectrum. Following the idea in Section~\ref{sec:3.3}, we modify the classical halo model to include the contribution from the unresolved mass, and compute the matter-matter power spectrum as detailed in Appendix~\ref{app:1}. One can see that the DHM achieves 5$\%$ accuracy in a $k$ range that covers the transition of 1-halo and 2-halo terms, by directly summing up the various components. 

\begin{figure}
	\includegraphics[width=\columnwidth]{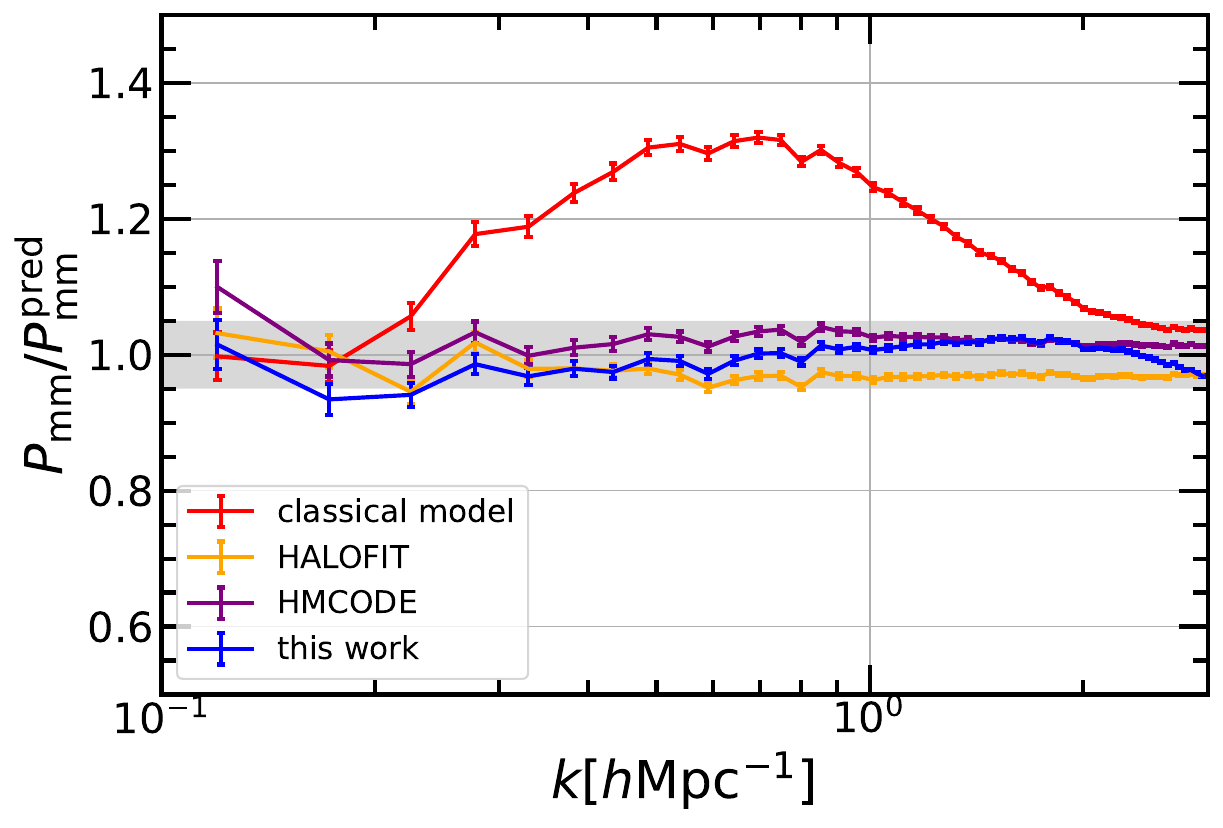}
    \caption{Ratios between the simulated and predicted power spectra using different models. Red curve is computed using virial-truncated profiles and a linearly-biased 2-halo term, as detailed in Section~\ref{sec:4.3}. Orange, purple and blue curves are computed using \textsc{halofit}~\citep[the version of][]{HaloFit12}, \textsc{hmcode}, and DHM. Error bars represent the standard errors estimated using Equation~\eqref{eq:P_errors}. The shaded areas mark the fractional errors less than 0.05.}
    \label{fig:PmmPred_comps}
\end{figure}

For comparison, Figure~\ref{fig:PmmPred_comps} shows the results from three other models. First, we construct a ``classical'' halo model adopting commonly-used recipes for its components, including a 2-halo term in proportion to the linear power spectrum, a halo mass function fit with \citet{ST02}, and an Einasto profile truncated at the virial radius with profile parameters specified in \citet{diemer2019accurate} and \citet{gao2008redshift}. We also compare the DHM with some fine-tuned models for predicting the power spectrum, including \textsc{halofit}~\citep{HaloFit03,HaloFit12} and \textsc{hmcode}~\citep{HMCODE15, HMCODE16, HMCODE21}. 

Focusing on the classical model, we find it performs poorly in this region. This under-prediction suggests a defect in this model, which is caused by the virial-truncated profile and a simplistic 2-halo term. According to the results in section~\ref{sec:4}, a matching profile is extended and have a smooth truncation. Using a virial-truncated profile will miss the contribution of the halo outskirts, leading to the underprediction of the matter-matter power spectrum. In addition, the 2-halo term also deviates from the linearly-biased linear power on small scale. On the other hand, DHM not only uses an extended Einasto profile to accurately model the mass distribution in the halo outskirts, but also carefully considers the halo exclusion and unresolved mass in the 2-halo term, improving the accuracy on transition scales. 

Compared with \textsc{halofit} and \textsc{hmcode}, DHM reaches a comparable accuracy. However, it is important to realize that the former two models are ``implicit" or pseudo halo models, while DHM is an explicit and physical halo model. In DHM, a halo population based on a physical boundary definition is explicitly identified and modeled, which facilitates further studies on the corresponding halos. By contrast, \textsc{halofit} is a fitting method without involving any halo property, and \textsc{hmcode} pertains to a population of ``effective halos" whose properties differ from those measured in simulations. Moreover, DHM is more interpretable. The model parameters in DHM have clear physical meanings and can all be measured from the halo population except for one parameter of the unresolved component which can still be derived from first principle. On the other hand, the parameters in \textsc{halofit} and \textsc{hmcode} are less interpretable because they do not correspond to any physical halo population. Finally, the physical nature of DHM enables it to predict multiple statistics simultaneously, including (but not limited to) the total matter and halo-matter power spectra as we have shown. By contrast, it is challenging to extend \textsc{halofit} and \textsc{hmcode} to predict the power spectra of other tracers because they use some ad-hoc fixes in modeling the matter-matter power spectrum. 




\begin{figure}
	\includegraphics[width=\columnwidth]{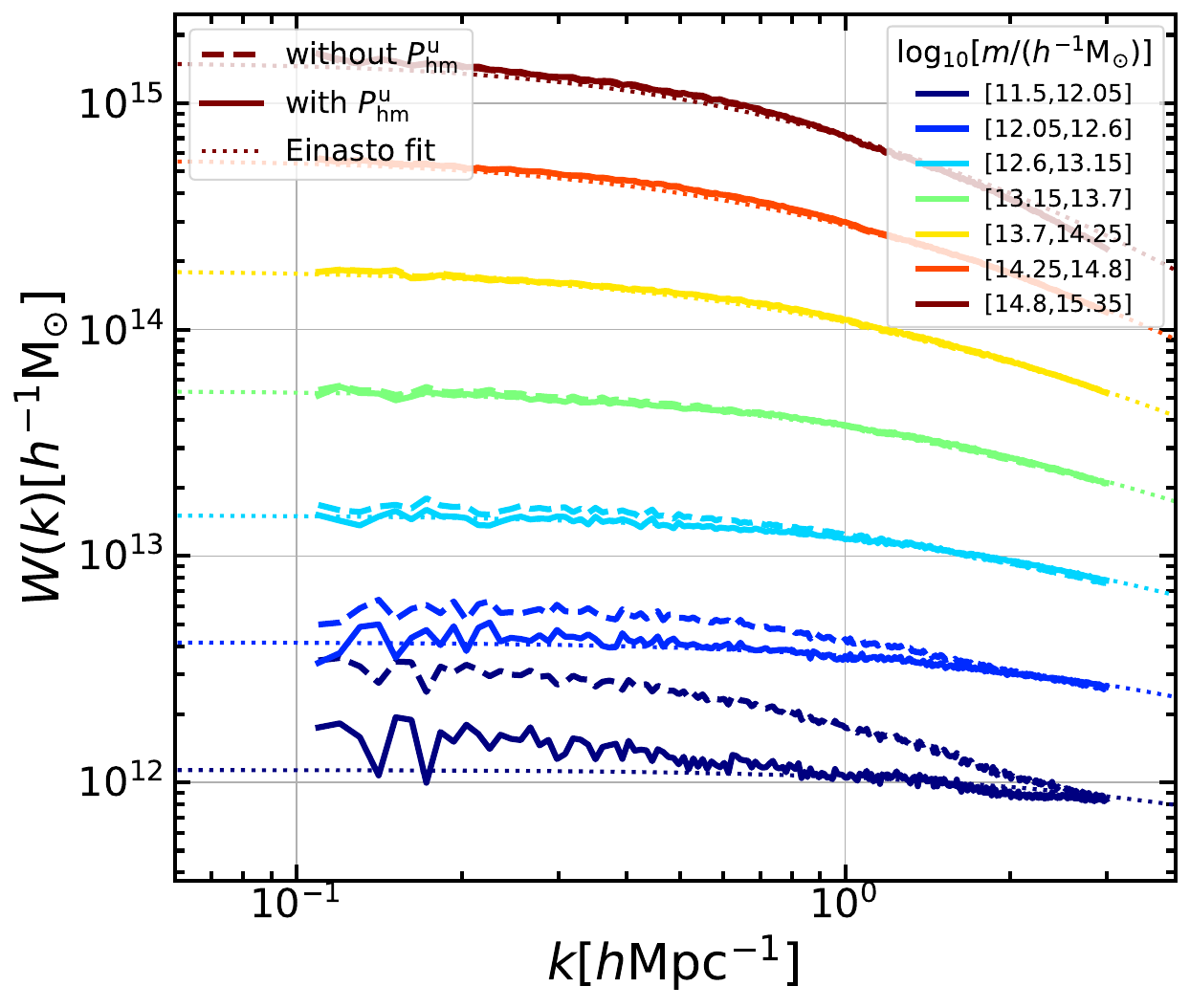}
    \caption{Same as Figure~\ref{fig:profiles_kspace}, but comparing the profile solutions of the $\Rid$ catalog with and without the unresolved term, $P^{\rm u}_{\rm hm}(k)$, in Equation~\eqref{eq:equations_DHM}, as labeled. }
    \label{fig:Profiles_nounr}
\end{figure}
\begin{figure*}
	\includegraphics[width=\textwidth]{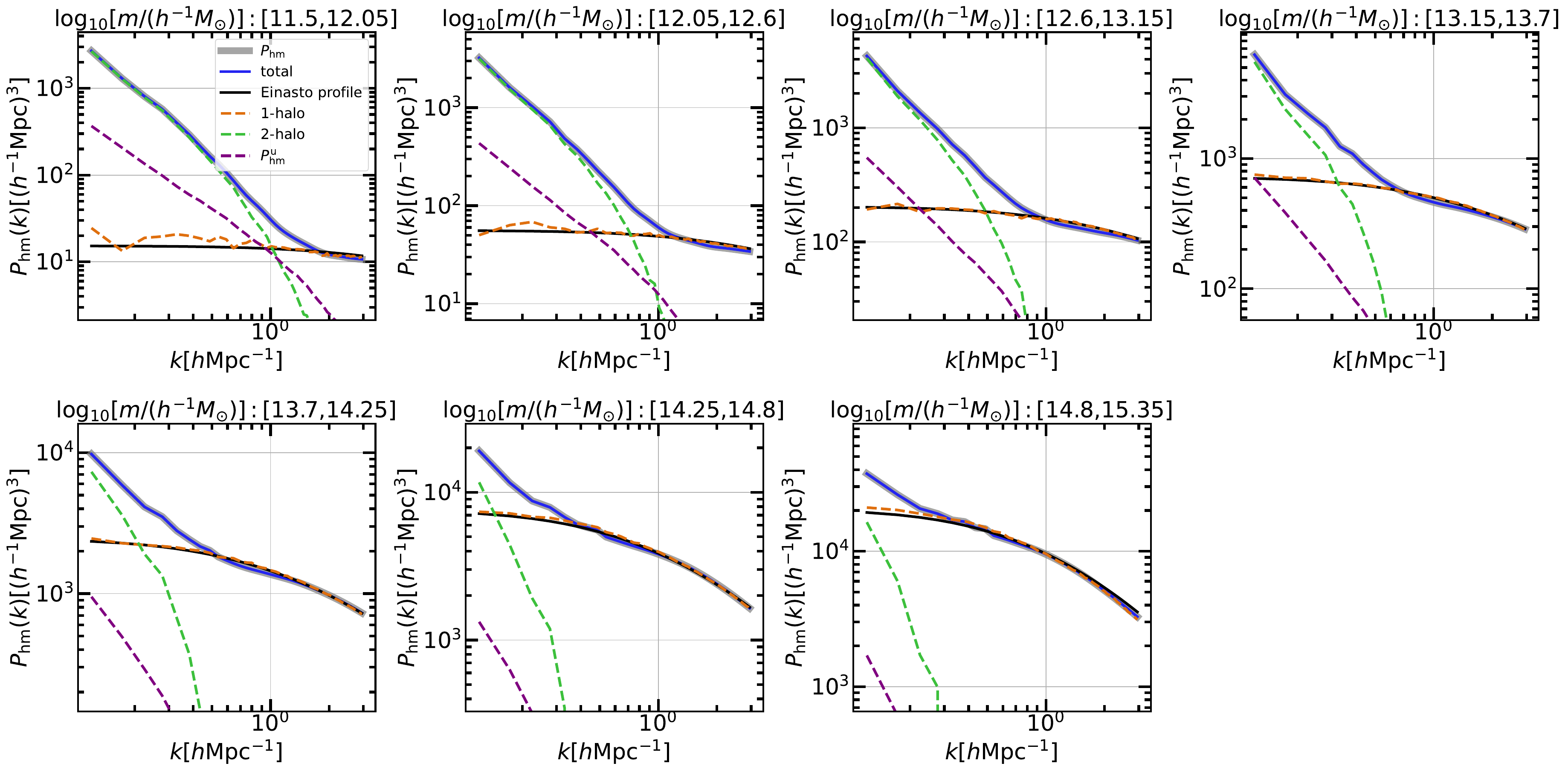}
    \caption{Decompositions of the halo-matter power spectra for the $\Rid$ catalog. In each panel, the thick grey curve shows the measured power spectrum for a given mass bin as labeled, while the blue solid curve shows the model prediction, with dashed curves showing the contributions from the 1-halo, 2-halo and unresolved terms respectively. The black solid curve shows the Einasto model as in Figure~\ref{fig:profiles_kspace}.}
    \label{fig:decomposition}
\end{figure*}

\section{Discussion} \label{sec:discussion}
\subsection{Profile uncertainty due to the unresolved term}


The modeling of the unresolved term relies on an extrapolation of the halo-halo correlation to the diffuse mass limit, in addition to a model for the integrated mass over the halo profile before solving for the matching profile. In this work, we have modeled these components following the treatment of \citet{DHM}, which is consistent with the depletion catalog but may not work for the other catalogs.

To show the influence of this modeling uncertainty, in Figure~\ref{fig:Profiles_nounr} we compare the solutions to Equation~\eqref{eq:equations_DHM} with and without the unresolved term for the depletion catalog. The solutions are barely changed for the high mass bins, while that for the lowest mass bin is significantly affected. This is not difficult to understand as the unresolved term should be most degenerate with halos close to being unresolved. Without the unresolved term, the profile of the lowest mass bin is significantly overestimated, to compensate for the missing mass and power. Similar overestimation of the matching profile can be observed in Figure~\ref{fig:profiles_kspace} for the lowest mass bin in the $\Rvir$ catalog, due to an underestimated unresolved term as detailed in Appendix~\ref{app:unresolved_term}.

The significance of the unresolved term can also be understood from its relative contribution to the power spectrum in each halo mass bin, as shown in Figure~\ref{fig:decomposition}. For the highest mass bin, the power spectrum is dominated by the 1-halo term over the entire $k$-range covered. For lower mass halos, the 2-halo term always dominates on the large scale. As the halo mass decrease, the contribution from the unresolved term increases, while the 1-halo term becomes less and less important. It is thus natural to expect that a variation in the unresolved term will have a significant influence on the halo profile of the low mass halos, while it barely affects that of high mass halos.

\subsection{The Navarro-Frenk-White (NFW) profile}
\begin{figure}
	\includegraphics[width=0.9\columnwidth]{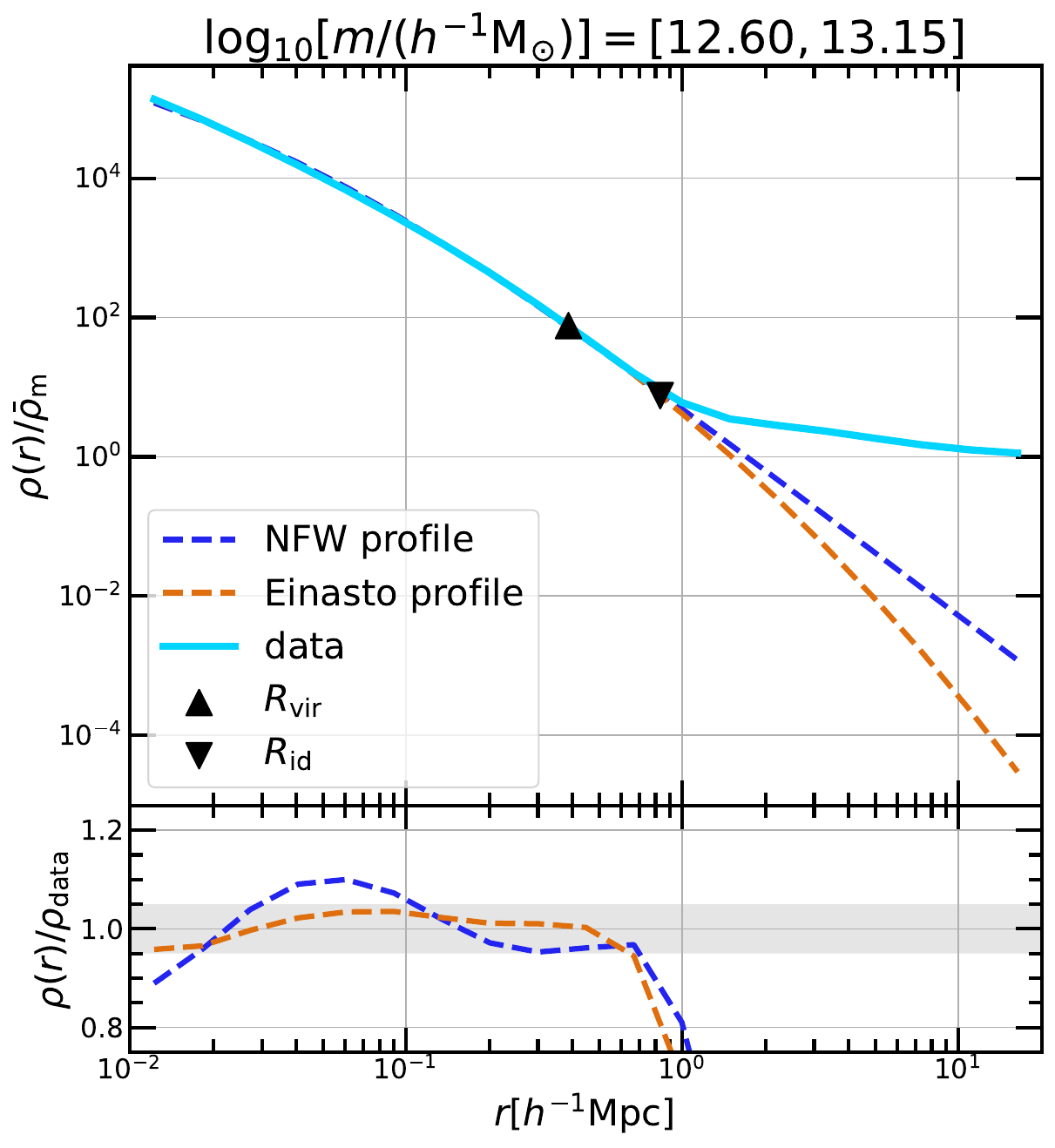}
    \caption{{Comparison of the NFW and Einasto profiles in configuration space. \textit{Top}: the solid curve is the stacked density profile around halos with mass $10^{12.60}<M/(\msunh)<10^{13.15}$. The blue and orange dashed curves are the NFW and Einasto fits, respectively. Their parameters are fitted using the data within $\Rid$. The upper and lower triangles mark the location of $\Rvir$ and $\Rid$. \textit{Bottom}: the ratio between the NFW and Einasto fits and stacked density profile. The shaded area marks the 5 percent variation range around unity.}}
    \label{fig:NFWvsEIN_rspace}
\end{figure}

\begin{figure*}
	\includegraphics[width=\textwidth]{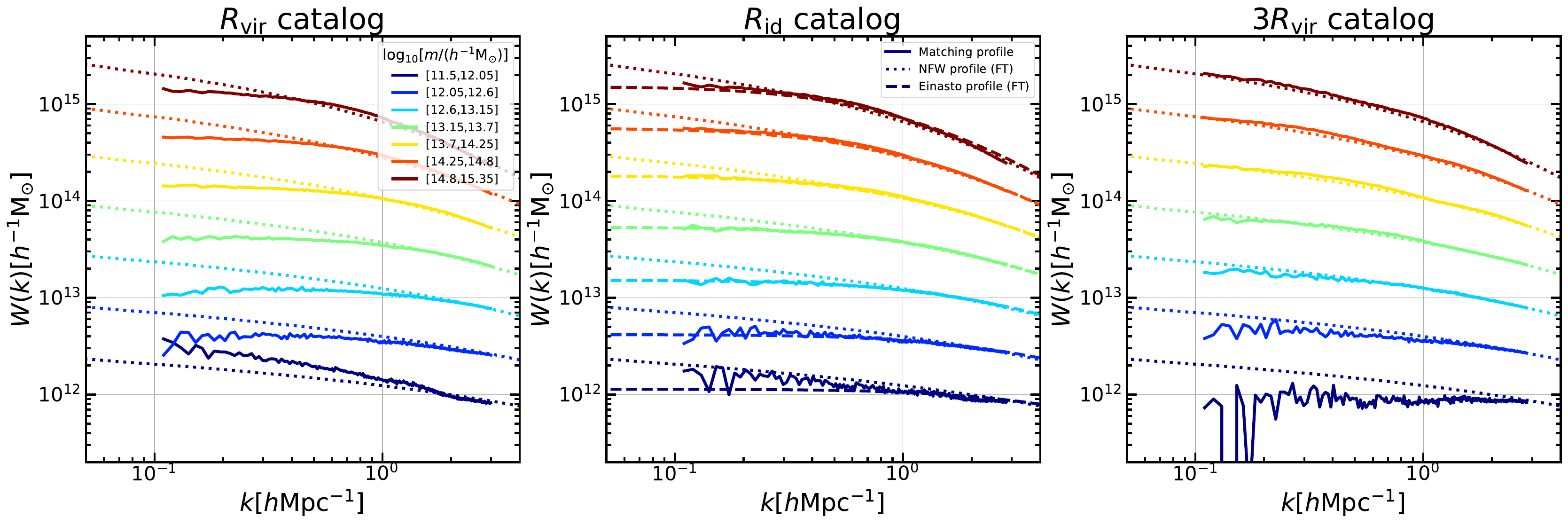}
    \caption{{The comparison of matching profiles and NFW fits. The solid (middle panel) and dashed (all panels) curves are the matching profiles and Einasto fits (identical with those in Figure~\ref{fig:profiles_kspace}). The dashed curves are the NFW fits. The parameters in NFW and Einasto profiles are obtained by fitting the stacked density profile within $\Rid$ in real space. For clarity, we only plot the Einasto profiles in the middle panel.}}
    \label{fig:profiles_NFW_kspace}
\end{figure*}

\begin{figure*}
	\includegraphics[width=\textwidth]{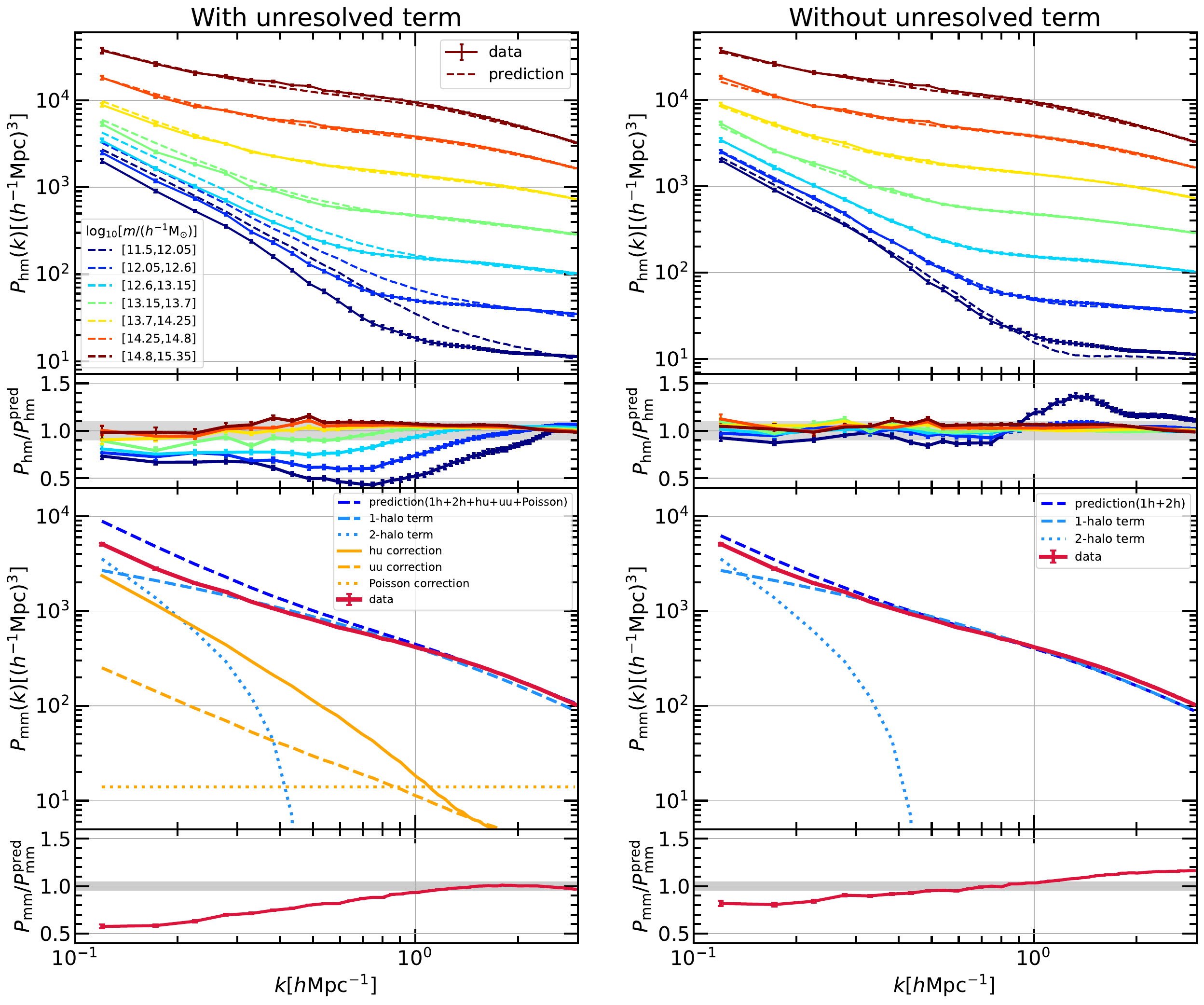}
    \caption{{The Power spectra predicted from the NFW-$3\Rvir$ scheme. \textit{Left}: The simulated power spectra, model prediction, and their ratios. The halo profiles are analytically modeled with the NFW profile. The halo bias and the halo mass function for the 3$\Rvir$ catalog are refit with the functional forms in Appendix~\ref{app:ingredients}, which differs from those for the $\Rid$ catalog. Analytically modeling of the unresolved term is detailed in Appendix~\ref{app:unresolved_term}. \textit{Right}: same as the left, but eliminating the contribution of unresolved mass. We set $P^{\rm u}_{\rm hm}(k|m)$ in Section~\ref{sec:3}, the halo-unresolved, unresolved-unresolved, and the Poisson terms in Appendix~\ref{app:1} equal to zero.}}
    \label{fig:PsPred_NFW}
\end{figure*}

In this section, we investigate how the Navarro-Frenk-White~\citep[NFW,][]{navarro1995simulations,navarro1996structure,navarro1997universal} profile compares with the matching profiles and the performance of a halo model built with it. The NFW profile is written as 
\begin{equation}
    \rho_{\rm NFW} = \frac{\rho_{\rm s}}{(r/r_{\rm s})(1+r/r_{\rm s})^2}.
\end{equation}
It involves two parameters,  the scale radius $r_{\rm s}$ and the normalization $\rho_{\rm s}$, which is simpler than the three-parameter Einasto profile. Some recent studies have shown that the Einasto profile is more accurate in describing the halo profile and is more universal across cosmologies \citep{Navarro10TheDiversity, Ludlow11Thedensity, brown20connecting, Wang20Universal}. Figure~\ref{fig:NFWvsEIN_rspace} provides a direct comparison of the NFW and Einasto profiles in fitting the average profile of a group-size halo. In agreement with previous results, the Einasto profile performs better within $\Rid$. Beyond $\Rid$,  the Einasto profile drops faster than the NFW profile, while the measured halo profile deviates from both NFW and Eiansto profiles due to contamination from masses associated with other halos.

Figure~\ref{fig:profiles_NFW_kspace} compares the best-fitting NFW profiles in each bin with the matching profiles in Fourier space. As shown in the middle panel, the NFW and Einasto profiles are similar on small scales and deviate from each other on large scales, consistent with Figure~\ref{fig:NFWvsEIN_rspace}. In addition, as the NFW profile has a diverging total mass, it keeps increasing with $k$ decreasing, while the Einasto profile converges to a finite mass as $k\rightarrow 0$. In all three panels, we compare the NFW profile with the matching profiles of the three catalogs. In the $\Rvir$ and $\Rid$ cases, the NFW profiles are generally higher than the matching profiles. Interestingly, for the $3\Rvir$ case, we find that the NFW profiles are higher than the matching profiles on large scales in low-mass bins (deep blue, blue, and light blue curves) but agree with the matching profiles in high-mass bins (yellow, orange, and brown). The high-mass agreements lead to a question that, can we also analytically model the halo-matter and matter-matter statistics by coupling the NFW profiles with the $3\Rvir$ catalog?

To answer this question, we first calibrate the model ingredients for the $3\Rvir$ catalog. We set the mass range of resolved halos as $M_{\rm min}=10^6\msunh$ to $M_{\rm max}=10^{16}\msunh$. For the halo bias and the halo mass function, we refit the parameters with the fitting functions provided in \citet{jing1998accurate} and \citet{ST02} (see details of the fitting functions in Appendix~\ref{app:ingredients}). The functional forms used in the NFW-$3\Rvir$ scheme are identical with those in the Einasto-$\Rid$ scheme, but with different best-fitting parameters. For the halo profile, we use the NFW profile with the mass-concentration relation provided in \citet{diemer2019accurate}. 

With all the ingredients above and the unresolved term modeled in Appendix~\ref{app:unresolved_term}, we obtain the predictions based on the NFW-$3\Rvir$ combination, shown in the left panels of Figure~\ref{fig:PsPred_NFW}. There are significant over-predictions in the low-mass $P_{\rm hm}$s and $P_{\rm mm}$ on large scales, which are likely caused by the over-predictions of low-mass profiles. On large scale ($k<1\kmpch$), NFW profiles are higher than the matching profiles for low mass halos in the $3\Rvir$ catalog, as shown in last panel of Figure~\ref{fig:profiles_NFW_kspace}. For high-mass $P_{\rm hm}$s, the NFW-$3\Rvir$ scheme performs well. The over-predictions of low-mass profiles do not affect the high-mass results significantly, because the high-mass $P_{\rm hm}$s is dominant by the 1-halo term and the 2-halo term contributes little in the $k$ range we focus on.

Note that in Equation~\eqref{eq:effective_bias} we estimate the effective bias of the unresolved mass $b_{\rm unr}$ using the integrated Einasto mass $M_{\rm EIN} = \int_{0}^{\infty}\rho_{\rm EIN}(\textbf{\textit{r}})\ud^3\textbf{\textit{r}}$, where $m_{\rm res}$ and $m_{\rm max}$ are the upper and lower mass limit of resolved halos. As the NFW profile integrates to larger (infinite in fact) mass than the Einasto profile, the adopted $b_{\rm unr}$ is overestimated for the NFW-$3\Rvir$ scheme. To test the influence of the unresolved term, in the right panels of Figure~\ref{fig:PsPred_NFW} we completely eliminate the unresolved components in the model. 
We find $P_{\rm hm}$s can be reproduced well with $10\%$ accuracy for $M>10^{12}\msunh$ in this case, while $P_{\rm mm}$ are still over-predicted on large scale. 
In addition, the lowest-mass $P_{\rm hm}$ and $P_{\rm mm}$ are under-estimated on small scales, which is due to the lack of the unresolved term.

These results are all rooted in the fact that the NFW profile is an extremely extended profile with an infinite integrated mass. To match such an extended profile, it is necessary to define halos according to a much larger boundary (3$R_{\rm vir}$ or larger) than conventionally done. Even so, the agreement cannot be perfect as the halo model solution requires a finite mass for each halo. As a result, a halo model adopting the NFW profile and the 3$R_{\rm vir}$ boundary still over predicts the large-scale power spectra. To improve the model prediction using the NFW profile, it then becomes necessary to truncate the NFW profile on a certain scale while defining the halo catalog accordingly, or artificially suppress the population of low mass halos or diffuse matter at the price of breaking self-consistency.


\section{Summary and conclusions} \label{sec:6}

In this work, we solve for the halo density profiles in Fourier space from measurements of the halo-matter and halo-halo power spectra in a cosmological simulation. This is possible because the matter field can be modeled by the halo field convolved with the halo density profiles. The Fourier space solution thus avoids ambiguities in partitioning matter among neighboring halos, enabling us to derive the complete halo profile out to large scale. 

We have applied the method to three halo catalogs with different boundary definitions, including the virial radius, $\Rvir$, the depletion radius, $\Rid$, and $3\Rvir$, to study the relations between halo profile and boundary definition, and their implications for halo model. Our main findings are as follows.



\begin{itemize}
    \item For each of the halo catalogs, a set of matching halo profiles can always be found to accurately reconstruct the input halo-matter power spectra. This implies there can be multiple ways of decomposing the matter field into halos, and the matter distribution around the halos can always be well described with a halo model, at least in terms of the halo-matter power spectrum.
    \item The matching profiles vary in different halo catalogs due to different population properties. Halos defined with a more extended boundary also have more extended outer profiles and larger integrated masses over the profiles. The profiles all extend beyond the exclusion radius used to define the halo catalogs, reflecting ongoing mass accretion and mergers which extend the mass distribution around halos.
    
    \item The matching profiles are nearly identical within $\Rid$ across the three catalogs. This indicates that the depletion region characterized by $\Rid$ is a universal feature around halos in the mass range covered.

    \item For the $\Rid$ catalog, the matching profiles are well described by the Einasto profiles over the scale covered by our measurements. This supports the choice of the Einasto profile in the depletion halo model of \citet{DHM}. For the $\Rvir$ catalog, the matching profiles drop faster than the Einasto profile outside the depletion radius.
    \item Proper modeling of unresolved halos and diffuse matter is important for extracting the profiles around low mass halos. Ignoring or underestimating the unresolved term can result in overestimation of the low mass profiles. 
    \item {The matching profiles for halos defined with a $3R_{\rm vir}$ boundary are close to the NFW profile over the finite $k$ scale investigated, especially for high mass halos. However, a halo model built from the NFW profile over-predicts the large-scale power spectra due to the diverging large-scale power of the NFW profile. To use NFW profile for power spectra prediction, it is necessary to define halos on a large boundary scale and truncate the NFW profile accordingly, or artificially suppress the contribution from low mass halos and diffuse matter.} 
\end{itemize}

The agreement between the numerical solution and the Einasto profile model for the depletion catalog enables us to construct an analytical halo model for this catalog, which is the Fourier space equivalence to the depletion halo model in \citet{DHM}. This model can predict the halo-matter power spectra to 10\% accuracy across scales and halo masses. Moreover, the same model also predicts the matter-matter power spectrum to 5\% accuracy. This illustrates the advantage of an explicit and physical halo model, that the same model can simultaneously predict multiple statistics of the halo and matter fields. For comparison, existing halo models for predicting the matter power spectrum such as \textsc{halofit} and \textsc{hmcode} can be regarded as implicit or effective models. These models do not directly correspond to an identifiable halo population, and therefore the model ingredients can not be directly verified with simulations. As a result, the predicting power of these implicit models is usually limited to the total matter power spectrum which is used for training the model.


In this work, we have only constructed and verified the DHM for a single snapshot at $z=0$. In future works, we will extend our model to different redshifts and cosmologies, by studying the redshift evolution and cosmology dependence of the model ingredients. This may eventually lead to a precise, unified and physical understanding of structure formation using halos as building blocks.

Finally, we note that some recent works showed that the halo model sensitively responds to the effects of massive neutrinos~\citep[e.g.,][]{hannestad20spoon,cataneo20otrtp} and baryons~\citep[e.g.,][]{acuto21BAHAMAS, bose21otrtp} in the transition region. Given its ability to accurately model the matter distribution in this region, the DHM promises to provide a better understanding of these physical processes and model their effects accurately. We plan to address these problems in future works.

\appendix

\section{Analytical model ingredients in the Depletion-radius-based halo model}
\label{app:ingredients}
In this Appendix, we specify the ingredients when implementing the Depletion-radius-based halo model, including the analytical modeling of the halo-halo correlation function, halo mass function, and halo profile according to \citet{DHM}.

The halo-halo correlation function can be modeled as,
\begin{equation}
    \xi_{\rm hh}(r|m_1,m_2) = b_{\rm hh}(m_1)b_{\rm hh}(m_2)\hat{\xi}_{\rm hh}(r)\mathrm{H}\{r-[R_{\rm id}(m_1)+R_{\rm id}(m_2)]\},
    \label{eq:CF_hh_DHM}
\end{equation}
where $\hat{\xi}_{\rm hh}(r)$ is the universal halo correlation that can be modeled by a power-law\begin{equation}
    \hat{\xi}_{\rm hh}(r)=\left(\frac{r}{r_0}\right)^{-\gamma}
    \label{eq:xi_hh_star}
\end{equation} with best-fitting parameters $r_0=4.96$ and $\gamma=1.58$ in the $\Rid$ catalog. 
The Heaviside step function, ${\rm H}(x)$, is unity at $x>0$ and 0 otherwise. The halo bias is fit using the formula suggested by \citet{jing1998accurate},
\begin{equation}
    b_{\rm hh}(M_{\rm vir}) = \left[\frac{0.5}{\nu^4(M_{\rm vir})}+1\right]^c[1+D\nu^d(M_{\rm vir})+E],
    \label{eq:bias_DHM}
\end{equation}
where $\nu(M_{\rm vir})=\delta_{\rm sc}/\sigma(M_{\rm vir})$ is the peak height, $\sigma(m)$ is the variance of the linear density field within a top-hat filter containing mass $M_{\rm vir}$, $\delta_{\rm sc}\approx 1.686$ is critical overdensity for collapse derived from the spherical collapse model. The best-fitting parameters $c=0.206$, $d=1.494$, $D=0.731$, $E=-0.959$. 

The hao mass function is fit with the Sheth $\&$ Tormen formula (\citealt{sheth1999large, sheth2001ellipsoidal,ST02}),
\begin{equation}
    f_{\rm ST}(\nu)=F\sqrt{\frac{2a}{\pi}}\nu [1+(a\nu^2)^{-p}]e^{-\frac{a\nu^2}{2}},
    \label{eq:MF_DHM}
\end{equation}
where $F=0.2677$, $a=0.7765$, $p=-0.0115$ for the $\Rid$ catalog. The halo number density can be derived from above equation through
\begin{equation}
n(M_{\rm vir})\ud M_{\rm vir}=\frac{\bar{\rho}}{M_{\rm vir}}f_{\rm ST}(\nu)\frac{\ud \nu}{\nu},
    \label{eq:num_dens_DHM}
\end{equation}
where $\bar{\rho}$ is the mean matter density of the Universe, and the peak height $\nu(M_{\rm vir})$ corresponds to the virial mass $M_{\rm vir}$.

The halo density profile is modeled by the Einasto formula ~\citep{einasto1965construction, Merritt06, Navarro04, Navarro10}, 
\begin{equation}
    \rho_{\rm EIN}(r) = \rho_{\rm s} {\rm exp}\left( -\frac{2}{\alpha}\left[ \left(\frac{r}{r_{\rm s}}\right)^\alpha-1 \right]\right),
    \label{eq:EIN_DHM}
\end{equation}
with the mass-concentration suggested by \citet{diemer2019accurate} and $\nu-M$ relation suggested by \citet{gao2008redshift}.

\section{Modeling the unresolved term}
\label{app:unresolved_term}
According to \citet{DHM}, the halo-halo correlation functions follow a universal shape. The unresolved term can be modeled by generalizing the universal halo correlation down to the diffuse matter limit, expressed as
\begin{equation}
    \xi^{\rm unr}_{\rm hm}(r|M) = b_{\rm unr}b_{\rm hh}(M)\hat{\xi}_{\rm hh}(r)\mathrm{H}[r-R_{\rm id}(M)],
    \label{eq:CF_unr}
\end{equation}
where $\hat{\xi}_{\rm hh}(r)$ is the unit halo correlation detailed in Equation~\eqref{eq:xi_hh_star}, and ${\rm H}(x)$ is the Heaviside step function. The bias of the unresolved mass, $b_{\rm unr}$, also named as effective bias, can be estimated using local mass conservation, 
\begin{equation}
    b_{\rm unr} \approx 1-\frac{1}{\bar{\rho}_{\rm m}}\int^{m_{\rm max}}_{m_{\rm res}}M_{\rm int}n(m)b_{\rm hh}(m) \ud m,
    \label{eq:effective_bias}
\end{equation}
where $m_{\rm max}$ and $m_{\rm res}$ are the mass limits of resolved halos, and $M_{\rm int}=\int \rho(\textbf{\textit{r}})\ud^3\textbf{\textit{r}}$ is the mass integral of the halo profile, which is not necessarily equal to a general mass label $m$. Estimating $b_{\rm unr}$ according to Equation~\eqref{eq:CF_unr} requires knowing the halo profile a priori. For the $\Rid$ catalog, we have shown that the Einasto formula can model the halo profile up to the transition scale, so that $b_{\rm unr}$ of the $\Rid$ catalog can be directly estimated. For the $\Rvir$ and $3\Rvir$ catalogs, the Einasto formula performs poorly in modeling the outer profiles, as shown in Section~\ref{sec:4.2}. 
Despite this, we still adopt the integrated mass of the Einasto profile when evaluating Equation~\eqref{eq:CF_unr} for all three catalogs for simplicity.
Meanwhile, the halo bias and mass function are fit separately using the parametric functions in \citet{DHM} for each catalog, as these quantities vary in different catalogs. As a result, we expect $b_{\rm unr}$ to be underestimated for the $\Rvir$ catalog but overestimated for the $3\Rvir$ catalog, due to over/under estimation of the integrated mass in the two catalogs respectively.

\section{Fourier version of the depletion-radius-based halo model} \label{app:1}
To derive the matter-matter power spectrum with the DHM, we write the matter overdensity field $\delta_{\rm m}(\textit{\textbf{x}})$ as 
\begin{equation}
    \delta_{\rm m}(\textit{\textbf{x}})=\frac{1}{\bar{\rho}_{\rm m}}\sum_j n(m_j)\delta_{\rm h}(\textit{\textbf{x}}|m_j)\circledast \rho(\textit{\textbf{x}}|m_j)+\delta^{\rm u}_{\rm m}(\textit{\textbf{x}}),
    \label{eq:full_matterfield}
\end{equation}
where $n(m)$, $\delta_{\rm h}(\textit{\textbf{x}}|m)$, and $\rho(\textit{\textbf{x}}|m)$ are the mean number density, the overdensity field and the profile of a halo population with mass $m$, respectively, and $\circledast$ is the convolution operation. The term $ \delta^{\rm u}_{\rm m}(\textit{\textbf{x}})$ represents the contribution of the unresolved mass. 
Considering the Fourier transform of Equation~\eqref{eq:full_matterfield}, we have
\begin{align}
    \delta_{\rm m}(\textit{\textbf{k}})=\frac{1}{\bar{\rho}_{\rm m}}\sum_j n(m_j)\delta_{\rm h}(\textit{\textbf{k}}|m_j)W(\textit{\textbf{k}}|m_j)+\delta^{\rm u}_{\rm m}(\textit{\textbf{k}}).
\end{align}
The matter-matter power spectrum is 
\begin{align}
    P_{\rm mm}(\textit{\textbf{k}}) &= \frac{1}{\bar{\rho}_{\rm m}^2}\sum_i\sum_j n(m_i)n(m_j)W(\textit{\textbf{k}}|m_i)W(\textit{\textbf{k}}|m_j)P_{\rm hh}(\textit{\textbf{k}}|m_i,m_j) \nonumber\\
    &+ \frac{2}{\bar{\rho}_{\rm m}}\sum_j n(m_j)W(\textit{\textbf{k}}|m_j)P_{\rm hu}(\textit{\textbf{k}}|m_j) \nonumber\\
    &+ P_{\rm uu}(\textit{\textbf{k}}),
    \label{eq:full_powerspectrum}
\end{align}
where $P_{\rm hu}(\textit{\textbf{k}}|m_j)$ and $P_{\rm uu}(\textit{\textbf{k}})$ are the cross power spectra of halo-unresolved and unresolved-unresolved.

The 1-halo term arises from $P_{\rm hh}(\textit{\textbf{k}}|m_j)$ when two halos are identical. For unresolved low mass halos, we can approximate them as point masses with $W(k|m\rightarrow0)\approx M_{\rm int}$. We thus break the 1-halo term into resolved and unresolved halos as
\begin{equation}
    P^{\rm 1h}_{\rm mm}(k) = \frac{1}{\bar{\rho}_{\rm m}^2}\sum_j^{\rm resolved} n(m_j)W^2(k|m_j) + \frac{1}{\bar{\rho}_{\rm m}^2}\int_{m_{\rm fs}}^{m_{\rm res}} M_{\rm int}^2 n(m)\ud m,
    \label{eq:full_1halo}
\end{equation} where $m_{\rm fs}$ and $m_{\rm res}$ are the freestreaming mass and the minimum mass of resolved halos considered by the model. The second term represents the shot-noise from point-mass halos. For the power spectrum measured from simulations, the diffuse matter are resolved by discrete particles and thus should also contribute a shot noise term. In the following we will not distinguish the two contributions but model them with a single constant $C$.

The 2-halo term involves contributions from the halo-halo, halo-unresolved, and unresolved-unresolved terms, expressed as
\begin{align}
    P^{\rm 2h}_{\rm mm}(k) &= \frac{1}{\bar{\rho}_{\rm m}^2}\sum_i\sum_{j\neq i} n(m_i)n(m_j)W(k|m_i)W(k|m_j)P^{\rm 2h}_{\rm hh}(k|m_i,m_j) \nonumber \\
    &+ \frac{2}{\bar{\rho}_{\rm m}}\sum_jn(m_j)W(k|m_i)P^{\rm 2h}_{\rm hu}(k|m_j) \nonumber \\
    &+ P^{\rm 2h}_{\rm uu}(k).
\end{align}
Note here the unresolved terms are contributed by both unresolved halos and diffuse matter.
 The power spectra $P^{\rm 2h}_{\rm hu}(k|m_j)$ and $P^{\rm 2h}_{\rm uu}(k)$ can be obtained from Fourier transforms of the correlation functions specified in \citet{DHM}, 
 \begin{align}
     \xi^{\rm 2h}_{\rm hu}(r|m_j) &= b_{\rm unr}b(m_j)\hat{\xi}_{\rm hh}(r){\rm H}[r-R_{\rm ex}(m_j)], \\
     \xi^{\rm 2h}_{\rm uu}(r) &= b^2_{\rm unr}\hat{\xi}_{\rm hh}(r).
 \end{align}

Replacing the summations with integrations, the final expression of the matter-matter power spectrum in the depletion halo model is
\begin{align}
P_{\rm mm}(k) &= P^{\rm 1h}_{\rm mm}(k)+P^{\rm 2h}_{\rm mm}(k) \\
\label{eq:Pmm_1h}
P^{\rm 1h}_{\rm mm}(k) &= \frac{1}{\bar{\rho}_{\rm m}^2}\int_{m_{\rm res}}^{m_{\rm max}}n(m)W^2(k|m)\ud m + C \\
\label{eq:Pmm_2h}
P^{\rm 2h}_{\rm mm}(k) &= \frac{1}{\bar{\rho}_{\rm m}^2}\int_{m_{\rm res}}^{m_{\rm max}}\int_{m_{\rm res}}^{m_{\rm max}}n(m_1)n(m_2)W(k|m_1)W(k|m_2)P^{\rm 2h}_{\rm hh}(k|m_1,m_2)\ud m_1 \ud m_2 \nonumber \\
&+ \frac{2}{\bar{\rho}_{\rm m}}\int_{m_{\rm res}}^{m_{\rm max}}n(m)W(k|m)P^{\rm 2h}_{\rm hu}(k|m)\ud m \nonumber \\
&+ P^{\rm 2h}_{\rm uu}(k).
\end{align}
where $m_{\rm res}$ and $m_{\rm max}$ represent the mass limits of resolved halos. We refer to the constant $C$ in Equation~\eqref{eq:Pmm_1h} as the Poisson term, the second and the third rows in Equation~\eqref{eq:Pmm_2h} as the halo-unresolved and unresolved-unresolved terms, respectively.

An alternative derivation of the above equations can be obtained from Fourier transforming the matter-matter correlation,
\begin{equation}
    \xi_{\rm mm}(r) = \int f(m)\xi_{\rm hm}(r|m)\circledast u_{\rm h}(r|m)\ud m ,
\end{equation} where $\xi_{\rm hm}(r|m)$ is modeled in \citet{DHM} and the integration continues down to the diffuse matter limit.

In the above model, we have not considered the existence of subhalos inside each halo, which could contribute additional small-scale power to the 1-halo term on top of the smooth density profile~\citep{Pmm_substructure1, Pmm_substructure2,substructure18Rivero}. A complete treatment will have to involve the distribution and internal structure of subhalos. For scales much larger than the sizes of subhalos, however, we can approximate subhalos as point masses, so that the subhalo contribution can also be approximated as a Poisson term, which can be absorbed into the parameter $C$ above. 
As shown in Section~\ref{sec:4.3}, the above model performs well for $0.1<k<3\kmpch$ with 5$\%$ accuracy, with a best-fitting parameter $C=16.04$ for our adopted $m_{\rm res}=10^{6}h^{-1}{\rm M}_{\odot}$ and $m_{\rm max}=10^{16}h^{-1}{\rm M}_{\odot}$. Nevertheless, one should keep in mind that the matter-matter power spectrum does not converge to a constant with $k$ increasing, and a scale-dependent correction term is needed if a higher precision is required. We will extend our model with subhalo terms in future works.


\section*{Acknowledgements}
We thank Yipeng Jing for access to the CosmicGrowth simulation, and Ji Yao and Pengjie Zhang for helpful discussions. This work is supported by National Key R\&D Program of China (2023YFA1607800, 2023YFA1607801), 111 project (No.\ B20019), and the science research grants from the China Manned Space Project (No.\ CMS-CSST-2021-A03). We acknowledge the sponsorship from Yangyang Development Fund. The computation of this work is done on the \textsc{Gravity} supercomputer at the Department of Astronomy, Shanghai Jiao Tong University. 


\bibliography{ref}{}

\begin{thebibliography}{}
\expandafter\ifx\csname natexlab\endcsname\relax\def\natexlab#1{#1}\fi
\providecommand{\url}[1]{\href{#1}{#1}}
\providecommand{\dodoi}[1]{doi:~\href{http://doi.org/#1}{\nolinkurl{#1}}}
\providecommand{\doeprint}[1]{\href{http://ascl.net/#1}{\nolinkurl{http://ascl.net/#1}}}
\providecommand{\doarXiv}[1]{\href{https://arxiv.org/abs/#1}{\nolinkurl{https://arxiv.org/abs/#1}}}

\bibitem[{{Acuto} {et~al.}(2021){Acuto}, {McCarthy}, {Kwan}, {Salcido}, {Stafford}, \& {Font}}]{acuto21BAHAMAS}
{Acuto}, A., {McCarthy}, I.~G., {Kwan}, J., {et~al.} 2021, \mnras, 508, 3519, \dodoi{10.1093/mnras/stab2834}

\bibitem[{{Adhikari} {et~al.}(2014){Adhikari}, {Dalal}, \& {Chamberlain}}]{splashback14Adhikari}
{Adhikari}, S., {Dalal}, N., \& {Chamberlain}, R.~T. 2014, \jcap, 2014, 019, \dodoi{10.1088/1475-7516/2014/11/019}

\bibitem[{{Allgood} {et~al.}(2006){Allgood}, {Flores}, {Primack}, {Kravtsov}, {Wechsler}, {Faltenbacher}, \& {Bullock}}]{triaxial2006allgood}
{Allgood}, B., {Flores}, R.~A., {Primack}, J.~R., {et~al.} 2006, \mnras, 367, 1781, \dodoi{10.1111/j.1365-2966.2006.10094.x}

\bibitem[{{Asgari} {et~al.}(2023){Asgari}, {Mead}, \& {Heymans}}]{Asgari23}
{Asgari}, M., {Mead}, A.~J., \& {Heymans}, C. 2023, The Open Journal of Astrophysics, 6, 39, \dodoi{10.21105/astro.2303.08752}

\bibitem[{{Aung} {et~al.}(2021){Aung}, {Nagai}, {Rozo}, \& {Garc{\'\i}a}}]{Aung21}
{Aung}, H., {Nagai}, D., {Rozo}, E., \& {Garc{\'\i}a}, R. 2021, \mnras, 502, 1041, \dodoi{10.1093/mnras/staa3994}

\bibitem[{{Bah{\'e}} {et~al.}(2013){Bah{\'e}}, {McCarthy}, {Balogh}, \& {Font}}]{strip13Bahe}
{Bah{\'e}}, Y.~M., {McCarthy}, I.~G., {Balogh}, M.~L., \& {Font}, A.~S. 2013, \mnras, 430, 3017, \dodoi{10.1093/mnras/stt109}

\bibitem[{{Baugh} \& {Efstathiou}(1994)}]{aliasing1}
{Baugh}, C.~M., \& {Efstathiou}, G. 1994, \mnras, 270, 183, \dodoi{10.1093/mnras/270.1.183}

\bibitem[{{Behroozi} {et~al.}(2014){Behroozi}, {Wechsler}, {Lu}, {Hahn}, {Busha}, {Klypin}, \& {Primack}}]{strip14Behroozi}
{Behroozi}, P.~S., {Wechsler}, R.~H., {Lu}, Y., {et~al.} 2014, \apj, 787, 156, \dodoi{10.1088/0004-637X/787/2/156}

\bibitem[{{Bernardeau} {et~al.}(2002){Bernardeau}, {Colombi}, {Gazta{\~n}aga}, \& {Scoccimarro}}]{P_error2}
{Bernardeau}, F., {Colombi}, S., {Gazta{\~n}aga}, E., \& {Scoccimarro}, R. 2002, \physrep, 367, 1, \dodoi{10.1016/S0370-1573(02)00135-7}

\bibitem[{{Bertschinger}(1985)}]{Bertschinger85splash}
{Bertschinger}, E. 1985, \apjs, 58, 39, \dodoi{10.1086/191028}

\bibitem[{{Bose} {et~al.}(2021){Bose}, {Wright}, {Cataneo}, {Pourtsidou}, {Giocoli}, {Lombriser}, {McCarthy}, {Baldi}, {Pfeifer}, \& {Xia.}}]{bose21otrtp}
{Bose}, B., {Wright}, B.~S., {Cataneo}, M., {et~al.} 2021, \mnras, 508, 2479, \dodoi{10.1093/mnras/stab2731}

\bibitem[{{Bose} \& {Loeb}(2021)}]{Bose21}
{Bose}, S., \& {Loeb}, A. 2021, \apj, 912, 114, \dodoi{10.3847/1538-4357/abec77}

\bibitem[{{Brown} {et~al.}(2020){Brown}, {McCarthy}, {Diemer}, {Font}, {Stafford}, \& {Pfeifer}}]{brown20connecting}
{Brown}, S.~T., {McCarthy}, I.~G., {Diemer}, B., {et~al.} 2020, \mnras, 495, 4994, \dodoi{10.1093/mnras/staa1491}

\bibitem[{{Bryan} \& {Norman}(1998)}]{sphericalcollapse2}
{Bryan}, G.~L., \& {Norman}, M.~L. 1998, \apj, 495, 80, \dodoi{10.1086/305262}

\bibitem[{{Cataneo} {et~al.}(2020){Cataneo}, {Emberson}, {Inman}, {Harnois-D{\'e}raps}, \& {Heymans}}]{cataneo20otrtp}
{Cataneo}, M., {Emberson}, J.~D., {Inman}, D., {Harnois-D{\'e}raps}, J., \& {Heymans}, C. 2020, \mnras, 491, 3101, \dodoi{10.1093/mnras/stz3189}

\bibitem[{{Chen} \& {Afshordi}(2023)}]{ADM2}
{Chen}, A.~Y., \& {Afshordi}, N. 2023, \prd, 107, 103526, \dodoi{10.1103/PhysRevD.107.103526}

\bibitem[{{Cooray} \& {Sheth}(2002)}]{cooray2002halo}
{Cooray}, A., \& {Sheth}, R. 2002, \physrep, 372, 1, \dodoi{10.1016/S0370-1573(02)00276-4}

\bibitem[{{Cuesta} {et~al.}(2008){Cuesta}, {Prada}, {Klypin}, \& {Moles}}]{Cuesta08}
{Cuesta}, A.~J., {Prada}, F., {Klypin}, A., \& {Moles}, M. 2008, \mnras, 389, 385, \dodoi{10.1111/j.1365-2966.2008.13590.x}

\bibitem[{{Diaz Rivero} {et~al.}(2018){Diaz Rivero}, {Cyr-Racine}, \& {Dvorkin}}]{Pmm_substructure2}
{Diaz Rivero}, A., {Cyr-Racine}, F.-Y., \& {Dvorkin}, C. 2018, \prd, 97, 023001, \dodoi{10.1103/PhysRevD.97.023001}

\bibitem[{{D{\'\i}az Rivero} {et~al.}(2018){D{\'\i}az Rivero}, {Dvorkin}, {Cyr-Racine}, {Zavala}, \& {Vogelsberger}}]{substructure18Rivero}
{D{\'\i}az Rivero}, A., {Dvorkin}, C., {Cyr-Racine}, F.-Y., {Zavala}, J., \& {Vogelsberger}, M. 2018, \prd, 98, 103517, \dodoi{10.1103/PhysRevD.98.103517}

\bibitem[{{Diemer}(2018)}]{diemer2018colossus}
{Diemer}, B. 2018, \apjs, 239, 35, \dodoi{10.3847/1538-4365/aaee8c}

\bibitem[{{Diemer}(2022)}]{DiemerDynamical1}
---. 2022, \mnras, 513, 573, \dodoi{10.1093/mnras/stac878}

\bibitem[{{Diemer}(2023)}]{diemer2022dynamics}
---. 2023, \mnras, 519, 3292, \dodoi{10.1093/mnras/stac3778}

\bibitem[{{Diemer} \& {Joyce}(2019)}]{diemer2019accurate}
{Diemer}, B., \& {Joyce}, M. 2019, \apj, 871, 168, \dodoi{10.3847/1538-4357/aafad6}

\bibitem[{{Diemer} \& {Kravtsov}(2014)}]{DK14}
{Diemer}, B., \& {Kravtsov}, A.~V. 2014, \apj, 789, 1, \dodoi{10.1088/0004-637X/789/1/1}

\bibitem[{{Einasto}(1965)}]{einasto1965construction}
{Einasto}, J. 1965, Trudy Astrofizicheskogo Instituta Alma-Ata, 5, 87

\bibitem[{{Fillmore} \& {Goldreich}(1984)}]{Fillmore84splash}
{Fillmore}, J.~A., \& {Goldreich}, P. 1984, \apj, 281, 1, \dodoi{10.1086/162070}

\bibitem[{{Fong} \& {Han}(2021)}]{depletion1}
{Fong}, M., \& {Han}, J. 2021, \mnras, 503, 4250, \dodoi{10.1093/mnras/stab259}

\bibitem[{{Fong} {et~al.}(2022){Fong}, {Han}, {Zhang}, {Yang}, {Gao}, {Wang}, {Li}, {Katsianis}, \& {Alonso}}]{fong2022first}
{Fong}, M., {Han}, J., {Zhang}, J., {et~al.} 2022, \mnras, 513, 4754, \dodoi{10.1093/mnras/stac1263}

\bibitem[{{Gao} {et~al.}(2023){Gao}, {Han}, {Fong}, {Jing}, \& {Li}}]{depletion2}
{Gao}, H., {Han}, J., {Fong}, M., {Jing}, Y.~P., \& {Li}, Z. 2023, \apj, 953, 37, \dodoi{10.3847/1538-4357/acdfcd}

\bibitem[{{Gao} {et~al.}(2008){Gao}, {Navarro}, {Cole}, {Frenk}, {White}, {Springel}, {Jenkins}, \& {Neto}}]{gao2008redshift}
{Gao}, L., {Navarro}, J.~F., {Cole}, S., {et~al.} 2008, \mnras, 387, 536, \dodoi{10.1111/j.1365-2966.2008.13277.x}

\bibitem[{{Garc{\'\i}a} \& {Rozo}(2019)}]{Garcia19exclusion}
{Garc{\'\i}a}, R., \& {Rozo}, E. 2019, \mnras, 489, 4170, \dodoi{10.1093/mnras/stz2458}

\bibitem[{{Garc{\'\i}a} {et~al.}(2021){Garc{\'\i}a}, {Rozo}, {Becker}, \& {More}}]{Garcia2}
{Garc{\'\i}a}, R., {Rozo}, E., {Becker}, M.~R., \& {More}, S. 2021, \mnras, 505, 1195, \dodoi{10.1093/mnras/stab1317}

\bibitem[{{Garc{\'\i}a} {et~al.}(2023){Garc{\'\i}a}, {Salazar}, {Rozo}, {Adhikari}, {Aung}, {Diemer}, {Nagai}, \& {Wolfe}}]{Garcia3}
{Garc{\'\i}a}, R., {Salazar}, E., {Rozo}, E., {et~al.} 2023, \mnras, 521, 2464, \dodoi{10.1093/mnras/stad660}

\bibitem[{{Green} {et~al.}(2005){Green}, {Hofmann}, \& {Schwarz}}]{Green05}
{Green}, A.~M., {Hofmann}, S., \& {Schwarz}, D.~J. 2005, \jcap, 2005, 003, \dodoi{10.1088/1475-7516/2005/08/003}

\bibitem[{{Gunn} \& {Gott}(1972)}]{gunn1972infall}
{Gunn}, J.~E., \& {Gott}, J.~Richard, I. 1972, \apj, 176, 1, \dodoi{10.1086/151605}

\bibitem[{{Hamilton}(1997)}]{P_error1}
{Hamilton}, A.~J.~S. 1997, \mnras, 289, 285, \dodoi{10.1093/mnras/289.2.285}

\bibitem[{{Han} {et~al.}(2018){Han}, {Cole}, {Frenk}, {Benitez-Llambay}, \& {Helly}}]{HBT}
{Han}, J., {Cole}, S., {Frenk}, C.~S., {Benitez-Llambay}, A., \& {Helly}, J. 2018, \mnras, 474, 604, \dodoi{10.1093/mnras/stx2792}

\bibitem[{{Han} {et~al.}(2012){Han}, {Jing}, {Wang}, \& {Wang}}]{han2012resolving}
{Han}, J., {Jing}, Y.~P., {Wang}, H., \& {Wang}, W. 2012, \mnras, 427, 2437, \dodoi{10.1111/j.1365-2966.2012.22111.x}

\bibitem[{{Hannestad} {et~al.}(2020){Hannestad}, {Upadhye}, \& {Wong}}]{hannestad20spoon}
{Hannestad}, S., {Upadhye}, A., \& {Wong}, Y. Y.~Y. 2020, \jcap, 2020, 062, \dodoi{10.1088/1475-7516/2020/11/062}

\bibitem[{{Hezaveh} {et~al.}(2016){Hezaveh}, {Dalal}, {Holder}, {Kisner}, {Kuhlen}, \& {Perreault Levasseur}}]{Pmm_substructure1}
{Hezaveh}, Y., {Dalal}, N., {Holder}, G., {et~al.} 2016, \jcap, 2016, 048, \dodoi{10.1088/1475-7516/2016/11/048}

\bibitem[{{Jing}(2019)}]{jing2019cosmicgrowth}
{Jing}, Y. 2019, Science China Physics, Mechanics, and Astronomy, 62, 19511, \dodoi{10.1007/s11433-018-9286-x}

\bibitem[{{Jing}(1998)}]{jing1998accurate}
{Jing}, Y.~P. 1998, \apjl, 503, L9, \dodoi{10.1086/311530}

\bibitem[{{Jing}(2005)}]{aliasing2}
---. 2005, \apj, 620, 559, \dodoi{10.1086/427087}

\bibitem[{{Jing} \& {Suto}(2002)}]{triaxial2002jing}
{Jing}, Y.~P., \& {Suto}, Y. 2002, \apj, 574, 538, \dodoi{10.1086/341065}

\bibitem[{{Li} \& {Han}(2021)}]{li2021outermost}
{Li}, Z.-Z., \& {Han}, J. 2021, \apjl, 915, L18, \dodoi{10.3847/2041-8213/ac0a7f}

\bibitem[{{Ludlow} {et~al.}(2009){Ludlow}, {Navarro}, {Springel}, {Jenkins}, {Frenk}, \& {Helmi}}]{Ludlow09subhalos}
{Ludlow}, A.~D., {Navarro}, J.~F., {Springel}, V., {et~al.} 2009, \apj, 692, 931, \dodoi{10.1088/0004-637X/692/1/931}

\bibitem[{{Ludlow} {et~al.}(2011){Ludlow}, {Navarro}, {White}, {Boylan-Kolchin}, {Springel}, {Jenkins}, \& {Frenk}}]{Ludlow11Thedensity}
{Ludlow}, A.~D., {Navarro}, J.~F., {White}, S. D.~M., {et~al.} 2011, \mnras, 415, 3895, \dodoi{10.1111/j.1365-2966.2011.19008.x}

\bibitem[{{Mansfield} {et~al.}(2017){Mansfield}, {Kravtsov}, \& {Diemer}}]{Mansfield17splash}
{Mansfield}, P., {Kravtsov}, A.~V., \& {Diemer}, B. 2017, \apj, 841, 34, \dodoi{10.3847/1538-4357/aa7047}

\bibitem[{{Mead} {et~al.}(2021){Mead}, {Brieden}, {Tr{\"o}ster}, \& {Heymans}}]{HMCODE21}
{Mead}, A.~J., {Brieden}, S., {Tr{\"o}ster}, T., \& {Heymans}, C. 2021, \mnras, 502, 1401, \dodoi{10.1093/mnras/stab082}

\bibitem[{{Mead} {et~al.}(2016){Mead}, {Heymans}, {Lombriser}, {Peacock}, {Steele}, \& {Winther}}]{HMCODE16}
{Mead}, A.~J., {Heymans}, C., {Lombriser}, L., {et~al.} 2016, \mnras, 459, 1468, \dodoi{10.1093/mnras/stw681}

\bibitem[{{Mead} {et~al.}(2015){Mead}, {Peacock}, {Heymans}, {Joudaki}, \& {Heavens}}]{HMCODE15}
{Mead}, A.~J., {Peacock}, J.~A., {Heymans}, C., {Joudaki}, S., \& {Heavens}, A.~F. 2015, \mnras, 454, 1958, \dodoi{10.1093/mnras/stv2036}

\bibitem[{{Merritt} {et~al.}(2006){Merritt}, {Graham}, {Moore}, {Diemand}, \& {Terzi{\'c}}}]{Merritt06}
{Merritt}, D., {Graham}, A.~W., {Moore}, B., {Diemand}, J., \& {Terzi{\'c}}, B. 2006, \aj, 132, 2685, \dodoi{10.1086/508988}

\bibitem[{{More} {et~al.}(2015){More}, {Diemer}, \& {Kravtsov}}]{more2015splashback}
{More}, S., {Diemer}, B., \& {Kravtsov}, A.~V. 2015, \apj, 810, 36, \dodoi{10.1088/0004-637X/810/1/36}

\bibitem[{{Navarro} {et~al.}(1995){Navarro}, {Frenk}, \& {White}}]{navarro1995simulations}
{Navarro}, J.~F., {Frenk}, C.~S., \& {White}, S. D.~M. 1995, \mnras, 275, 720, \dodoi{10.1093/mnras/275.3.720}

\bibitem[{{Navarro} {et~al.}(1996){Navarro}, {Frenk}, \& {White}}]{navarro1996structure}
---. 1996, \apj, 462, 563, \dodoi{10.1086/177173}

\bibitem[{{Navarro} {et~al.}(1997){Navarro}, {Frenk}, \& {White}}]{navarro1997universal}
---. 1997, \apj, 490, 493, \dodoi{10.1086/304888}

\bibitem[{{Navarro} {et~al.}(2004){Navarro}, {Hayashi}, {Power}, {Jenkins}, {Frenk}, {White}, {Springel}, {Stadel}, \& {Quinn}}]{Navarro04}
{Navarro}, J.~F., {Hayashi}, E., {Power}, C., {et~al.} 2004, \mnras, 349, 1039, \dodoi{10.1111/j.1365-2966.2004.07586.x}

\bibitem[{{Navarro} {et~al.}(2010{\natexlab{a}}){Navarro}, {Ludlow}, {Springel}, {Wang}, {Vogelsberger}, {White}, {Jenkins}, {Frenk}, \& {Helmi}}]{Navarro10}
{Navarro}, J.~F., {Ludlow}, A., {Springel}, V., {et~al.} 2010{\natexlab{a}}, \mnras, 402, 21, \dodoi{10.1111/j.1365-2966.2009.15878.x}

\bibitem[{{Navarro} {et~al.}(2010{\natexlab{b}}){Navarro}, {Ludlow}, {Springel}, {Wang}, {Vogelsberger}, {White}, {Jenkins}, {Frenk}, \& {Helmi}}]{Navarro10TheDiversity}
---. 2010{\natexlab{b}}, \mnras, 402, 21, \dodoi{10.1111/j.1365-2966.2009.15878.x}

\bibitem[{{Philcox} {et~al.}(2020){Philcox}, {Spergel}, \& {Villaescusa-Navarro}}]{Philcox20EHM}
{Philcox}, O. H.~E., {Spergel}, D.~N., \& {Villaescusa-Navarro}, F. 2020, \prd, 101, 123520, \dodoi{10.1103/PhysRevD.101.123520}

\bibitem[{{Pizzardo} {et~al.}(2023{\natexlab{a}}){Pizzardo}, {Geller}, {Kenyon}, {Damjanov}, \& {Diaferio}}]{Pizzardo23infalling1}
{Pizzardo}, M., {Geller}, M.~J., {Kenyon}, S.~J., {Damjanov}, I., \& {Diaferio}, A. 2023{\natexlab{a}}, \aap, 680, A48, \dodoi{10.1051/0004-6361/202347470}

\bibitem[{{Pizzardo} {et~al.}(2023{\natexlab{b}}){Pizzardo}, {Geller}, {Kenyon}, {Damjanov}, \& {Diaferio}}]{Pizzardo23infalling2}
---. 2023{\natexlab{b}}, \aap, 680, A48, \dodoi{10.1051/0004-6361/202347470}

\bibitem[{{Profumo} {et~al.}(2006){Profumo}, {Sigurdson}, \& {Kamionkowski}}]{Profumo06}
{Profumo}, S., {Sigurdson}, K., \& {Kamionkowski}, M. 2006, \prl, 97, 031301, \dodoi{10.1103/PhysRevLett.97.031301}

\bibitem[{{Salazar} {et~al.}(2024){Salazar}, {Rozo}, {Garc{\'\i}a}, {Kokron}, {Adhikari}, {Diemer}, \& {Osinga}}]{DynamicalHM24Salazar}
{Salazar}, E.~M., {Rozo}, E., {Garc{\'\i}a}, R., {et~al.} 2024, arXiv e-prints, arXiv:2406.04054, \dodoi{10.48550/arXiv.2406.04054}

\bibitem[{{Schneider} {et~al.}(2013){Schneider}, {Smith}, \& {Reed}}]{Schneider13}
{Schneider}, A., {Smith}, R.~E., \& {Reed}, D. 2013, \mnras, 433, 1573, \dodoi{10.1093/mnras/stt829}

\bibitem[{{Sheth} {et~al.}(2001){Sheth}, {Mo}, \& {Tormen}}]{sheth2001ellipsoidal}
{Sheth}, R.~K., {Mo}, H.~J., \& {Tormen}, G. 2001, \mnras, 323, 1, \dodoi{10.1046/j.1365-8711.2001.04006.x}

\bibitem[{{Sheth} \& {Tormen}(1999)}]{sheth1999large}
{Sheth}, R.~K., \& {Tormen}, G. 1999, \mnras, 308, 119, \dodoi{10.1046/j.1365-8711.1999.02692.x}

\bibitem[{{Sheth} \& {Tormen}(2002)}]{ST02}
---. 2002, \mnras, 329, 61, \dodoi{10.1046/j.1365-8711.2002.04950.x}

\bibitem[{{Shi}(2016)}]{Shi16Splash}
{Shi}, X. 2016, \mnras, 459, 3711, \dodoi{10.1093/mnras/stw925}

\bibitem[{{Smith} {et~al.}(2003){Smith}, {Peacock}, {Jenkins}, {White}, {Frenk}, {Pearce}, {Thomas}, {Efstathiou}, \& {Couchman}}]{HaloFit03}
{Smith}, R.~E., {Peacock}, J.~A., {Jenkins}, A., {et~al.} 2003, \mnras, 341, 1311, \dodoi{10.1046/j.1365-8711.2003.06503.x}

\bibitem[{{Takahashi} {et~al.}(2012){Takahashi}, {Sato}, {Nishimichi}, {Taruya}, \& {Oguri}}]{HaloFit12}
{Takahashi}, R., {Sato}, M., {Nishimichi}, T., {Taruya}, A., \& {Oguri}, M. 2012, \apj, 761, 152, \dodoi{10.1088/0004-637X/761/2/152}

\bibitem[{{Tinker} {et~al.}(2005){Tinker}, {Weinberg}, {Zheng}, \& {Zehavi}}]{tinker2005mass}
{Tinker}, J.~L., {Weinberg}, D.~H., {Zheng}, Z., \& {Zehavi}, I. 2005, \apj, 631, 41, \dodoi{10.1086/432084}

\bibitem[{{Tomooka} {et~al.}(2020){Tomooka}, {Rozo}, {Wagoner}, {Aung}, {Nagai}, \& {Safonova}}]{Tomooka20}
{Tomooka}, P., {Rozo}, E., {Wagoner}, E.~L., {et~al.} 2020, \mnras, 499, 1291, \dodoi{10.1093/mnras/staa2841}

\bibitem[{{van den Bosch} {et~al.}(2013){van den Bosch}, {More}, {Cacciato}, {Mo}, \& {Yang}}]{van2013cosmological}
{van den Bosch}, F.~C., {More}, S., {Cacciato}, M., {Mo}, H., \& {Yang}, X. 2013, \mnras, 430, 725, \dodoi{10.1093/mnras/sts006}

\bibitem[{{Villaescusa-Navarro}(2018)}]{Pylians}
{Villaescusa-Navarro}, F. 2018, {Pylians: Python libraries for the analysis of numerical simulations}, Astrophysics Source Code Library, record ascl:1811.008.
\newblock \doeprint{1811.008}

\bibitem[{{Wang} {et~al.}(2020){Wang}, {Bose}, {Frenk}, {Gao}, {Jenkins}, {Springel}, \& {White}}]{Wang20Universal}
{Wang}, J., {Bose}, S., {Frenk}, C.~S., {et~al.} 2020, \nat, 585, 39, \dodoi{10.1038/s41586-020-2642-9}

\bibitem[{{Wang} {et~al.}(2022){Wang}, {Wang}, \& {Mo}}]{anisotropy2022wang}
{Wang}, X., {Wang}, H., \& {Mo}, H.~J. 2022, \aap, 667, A99, \dodoi{10.1051/0004-6361/202244338}

\bibitem[{{Zhou} \& {Han}(2023)}]{DHM}
{Zhou}, Y., \& {Han}, J. 2023, \mnras, 525, 2489, \dodoi{10.1093/mnras/stad2375}

\end{thebibliography}
\bibliographystyle{aasjournal}



\end{document}